\documentclass[12pt]{article}

\usepackage[a4paper,lmargin=1in,rmargin=1in,top=0.71in,bottom=0.71in]{geometry} 
\usepackage{setspace} 
\usepackage{amsmath}
\usepackage{amssymb}
\usepackage{derivative}
\usepackage{epsfig}
\usepackage{epstopdf} 
\usepackage{combelow} 
\usepackage{multirow} 
\usepackage{braket}
\usepackage[
singlelinecheck=false 
]{caption} 

\usepackage{indentfirst} 
\usepackage{titlesec} 
\titlelabel{\thetitle.\quad}
\titleformat*{\section}{\normalsize\bfseries}
\titleformat*{\subsection}{\normalsize\bfseries}
\titleformat*{\subsubsection}{\normalsize\bfseries}

\usepackage{xcolor}
\usepackage{ulem}

\newcommand{\p}{\partial}
\newcommand{\para}[1]{\left(#1\right)}
\newcommand{\rpara}[1]{\left[#1\right]}

\usepackage{tensor}
\usepackage{tikz}
\usepackage[justification=centering]{caption}
\usepackage{hyperref}
\hypersetup{colorlinks=true, linkcolor=blue, urlcolor=blue, citecolor=red}
\usepackage{caption}
\usepackage{subcaption}
\usepackage{float}
\usepackage{url}

\begin{document}

\setcounter{page}{1}

\begin{center}
{\bf
\uppercase{
Numerical Simulation of Black Hole Images from photon trajectories in Schwarzschild Geometry.
}
}
\end{center}

\begin{center}
{

Alin Galea${}^{\rm a}$, Victor E. Ambru\cb{s}${}^{\rm a, \dagger}$

\

{
\footnotesize
{\it  ${}^{\rm a}$ Department of Physics, West University of Timi\cb{s}oara\\
  300223 V. P\^ arvan Ave 4, Timi\cb{s}oara, Romania}

${}^{\dagger}$ victor.ambrus@e-uvt.ro}
}
\end{center}

\

\

{
\footnotesize
With the arrival of advanced telescopes such as the Event Horizon Telescope (EHT), direct imaging of a black hole's shadow became a reality. Simulations play an important role in reproducing their appearance and constraining physical parameters of the black hole. We present a simple framework of simulating images of Schwarzschild black holes surrounded by thin, rotating accretion disks, accounting for gravitational and Doppler redshifts. A ray tracing algorithm based on integrating the photon geodesic equations with a 4th-order Runge-Kutta scheme is used for imaging. The accuracy of the numerical method is verified by convergence tests. A brief discussion of the validity of the weak field approximation is included and a qualitative comparison between the simulated images and the EHT image of M87* is presented.
}

\

\onehalfspacing

\section{Introduction}\label{sec:intro}

Ever since the observation of the M87* black hole by the Event Horizon Telescope (EHT) collaboration \cite{bib:EHT2019,EHT:2019uob}, black hole imaging has seen a huge spark in interest.

The subject of black holes has a long history, with the first black hole solution obtained in 1915 by Karl Schwarzschild by solving the vacuum Einstein equations \cite{bib:Hobson}. This solution characterizes a static, compact configuration of mass $M$, surrounded by an event horizon, located at the Schwarzschild radius $R_S = 2GM / c^2$. More elaborate models take into account a non-vanishing electric charge (the Reissner-Nordstr{\o}m solution) and/or black holes with non-vanishing angular momentum (the Kerr solution) -- see Ref.~\cite{bib:Hobson} for comprehensive discussions.

The study of black hole imaging has evolved over several decades, beginning with the pioneering works of Luminet \cite{Luminet:1979nyg} introducing the ray-tracing algorithm by solving the geodesic equation for photons on the Schwarzschild background geometry and advancing considerably to the achievements of the EHT collaboration \cite{bib:EHT2019,EHT:2024dhe}.

The image of a black hole is given by the accretion disk surrounding it, as the black hole itself does not emit radiation. As is well known, stable circular trajectories exist on the Schwarzschild geometry only down to a radius $R_{\rm ISCO} = 3R_S$, where the innermost stable circular orbit (ISCO) resides \cite{Shapiro:1983du}. The empty space between the ISCO and the black hole horizon, as well as the black hole itself, give rise to a dark patch in the black hole image, known as the black hole shadow \cite{Luminet:1979nyg,EHT:2024dhe,EHT:2025dua}.

Realistic models employed nowadays to intepret the telescope images rely on realistic general relativistic magnetohydrodynamics (GRMHD) solvers to describe the dynamics of the accretion disk, that evolve on realistic, Kerr-like background black hole spacetimes \cite{Porth:2016rfi,Prather:2024hsu}.

The aim of this work is to consider the problem of black hole imaging in the simplest setup: a static Schwarzschild background black hole spacetime, surrounded by a non-dynamical thin accretion disk. We develop a Runge-Kutta integration algorithm to compute the intersection between the photon trajectories and the accretion disk. Combining the gravitational redshift together with a simple model for the temperature distribution within the disk, we obtain images of static black holes surrounded by accretion disks of arbitrary inclination angle $i$. We also consider the Doppler shifts arising when the disk is rotating around the black hole according to the Kepler orbit formula. Our simplistic model reproduce with reasonable accuracy the image of the M87* black hole captured by the EHT collaboration \cite{bib:EHT2019,EHT:2019uob}.

This paper is structured as follows. For completeness and in order to establish our notation, we present in Sec.~\ref{sec:sch} a derivation of the Schwarzschild metric. In Sec.~\ref{sec:geo}, we derive the geodesic equation for photons on the Schwarschild background spacetime and compute the deflection angle in the weak-field approximation. Sec.~\ref{sec:num} introduces our numerical solver, validated against the weak-field solution. Sec.~\ref{sec:bh_image} illustrates the application of our algorithm for the imaging of black holes surrounded by thin accretion disks inclined with an angle $i$ with respect to the observer line. Subsec.~\ref{sec:bh_image:rot} considers the effect of rotation when the disk constituents follow the Kepler cicular orbit law, presenting a comparison with the image of the M87* black hole. Section~\ref{sec:conc} concludes this paper. Throughout this paper, we use the mostly-minus metric signature convention.

\section{Schwarzschild metric}\label{sec:sch}

Lensing effects can be most easily understood by studying the solution of Einstein field equations in the case of a spherically symmetric matter distribution. In this section, we revisit the derivation of the Schwarzschild solution, following Ref.~\cite{bib:Hobson}, introducing our notation.

We start from the general expression for the line element
\begin{equation}
    ds^2=g_{\mu\nu}dx^\mu dx^\nu,
\end{equation}

We wish to find a set of coordinates $\displaystyle x^\mu$ for which the metric is spherically symmetric (isotropic). In this case $ds^2$ must depend only on rotational invariants of spacelike coordinates $\displaystyle x^i$ and their differentials, which are
\begin{equation}
    \vec{x}\cdot\vec{x}\equiv r^2, \qquad d\vec{x}\cdot d\vec{x}, \qquad \vec{x}\cdot d\vec{x}. \notag
\end{equation}

The general form of a spatially isotropic line element is
\begin{equation}
    ds^2=A(t,r) c^2 dt^2 - B(t,r) c dt \hspace{3px}\vec{x}\cdot d\vec{x} - C(t,r)\para{\vec{x}\cdot d\vec{x}}^2 - D(t,r)d \vec{x}^2,
\end{equation}
where $A$, $B$, $C$ and $D$ are arbitrary functions of $t$ and $r$.
Transforming to spherical coordinates, the line element takes the form
\begin{equation}
    ds^2=A(t,r) c^2 dt^2-B(t,r)c dtdr-C(t,r)dr^2-D(t,r)\para{d\theta^2+\sin^2\theta \hspace{3px}d\phi^2}.
\end{equation}

We can introduce a new radial coordinate via $\overline{r}^2 = D(t,r)$, such that
\begin{equation}
    ds^2=A(t,\overline{r}) c^2 dt^2-B(t,\overline{r})c dtd\overline{r}-C(t,\overline{r})d\overline{r}^2-\overline{r}^2\para{d\theta^2+\sin^2\theta \hspace{3px}d\phi^2}.
\end{equation}
We also introduce a new temporal coordinate $\overline{t}$ defined by
\begin{equation}
    \label{newtimelike}
    cd\overline{t}=\Phi(t,\overline{r})\rpara{A(t,\overline{r})c dt-\frac{1}{2}B(t,\overline{r})d\overline{r}},
\end{equation}
where $\Phi(t,\overline{r})$ is an integrating factor that makes the right-hand side an exact differential.
By squaring (\ref{newtimelike}), we find the following
\begin{equation}
    A c^2 dt^2-Bc dt\hspace{3px}d\overline{r}=\frac{1}{A\Phi^2} c^2 d\overline{t}^2-\frac{B^2}{4A}d\overline{r}^2.
\end{equation}

Defining the new functions $\displaystyle \overline{A}=1/(A\Phi^2)$ and $\displaystyle \overline{B}=C+B^2/(4A)$, the metric becomes diagonal and the line elements takes the form
\begin{equation}
    ds^2=\overline{A}(\overline{t},\overline{r}) c^2 d\overline{t}^2 - \overline{B}(\overline{t},\overline{r})d\overline{r}^2-\overline{r}^2(d\theta^2+sin^2\theta \hspace{3px}d\phi^2).
\end{equation}
At last, we impose the condition that the metric is stationary to arrive at
\begin{equation}
    \label{general isotropic metric}
    ds^2=A(r)c^2 dt^2-B(r)dr^2-r^2(d\theta^2+sin^2\theta\hspace{3px}d\phi^2),
\end{equation}
where we dropped the bar notation.
The functions $A(r)$ and $B(r)$ are found by solving the Einstein field equations.

In our study, we are interested in the spacetime geometry outside of a spherical mass distribution (so $T_{\mu\nu}=0$). In this case, we must solve for the empty-space field equations which require that the components of the Ricci tensor must vanish
\begin{equation}
    R_{\mu\nu}=0.
\end{equation}
The expression of the Ricci tensor is:
\begin{equation}
    R_{\mu\nu}=\p_\nu\tensor{\Gamma}{^\sigma_\mu_\sigma}-\p_\sigma\tensor{\Gamma}{^\sigma_\mu_\nu}+\tensor{\Gamma}{^\rho_\mu_\sigma}\tensor{\Gamma}{^\sigma_\nu_\rho}-\tensor{\Gamma}{^\rho_\mu_\nu}\tensor{\Gamma}{^\sigma_\rho_\sigma}.
\end{equation}

From the expression of the line element \eqref{general isotropic metric}, we can write the components of the metric tensor and its inverse
\begin{equation}
    \hspace{-10pt} g_{\mu\nu}=\begin{pmatrix}
        A(r) & 0 & 0 & 0 \\
        0 & -B(r) & 0 & 0 \\
        0 & 0 & -r^2 & 0 \\
        0 & 0 & 0 & -r^2\sin^2\theta
    \end{pmatrix}, \quad
    g^{\mu\nu}=\begin{pmatrix}
        \frac{1}{A(r)} & 0 & 0 & 0 \\
        0 & -\frac{1}{B(r)} & 0 & 0 \\
        0 & 0 & -\frac{1}{r^2} & 0 \\
        0 & 0 & 0 & -\frac{1}{r^2\sin^2\theta}
    \end{pmatrix}.
\end{equation}

The Christoffel symbols are given by
\begin{equation}
    \tensor{\Gamma}{^\sigma_\mu_\nu}=\frac{1}{2}g^{\sigma\rho}\para{\p_\nu g_{\rho\mu}+\p_\mu g_{\nu\rho}-\p_\rho g_{\mu\nu}}.
\end{equation}
We obtain the following non-vanishing Christoffel symbols:
\begin{align}
\tensor{\Gamma}{^t_t_r} &= \frac{A'}{2A},
& \tensor{\Gamma}{^r_t_t} &= \frac{A'}{2B},
& \tensor{\Gamma}{^r_r_r} &= \frac{B'}{2B}, \notag \\
\tensor{\Gamma}{^r_\theta_\theta} &= -\frac{r}{B},
& \tensor{\Gamma}{^r_\varphi_\varphi} &= -\frac{r\sin^2\theta}{B},
& \tensor{\Gamma}{^\theta_r_\theta} &= \frac{1}{r}, \notag \\
\label{Christoffels_gen}
\tensor{\Gamma}{^\theta_\varphi_\varphi} &= -\sin\theta\cos\theta,
& \tensor{\Gamma}{^\varphi_r_\varphi} &= \frac{1}{r},
& \tensor{\Gamma}{^\varphi_\theta_\varphi} &= \cot\theta,
\end{align}
where the prime denotes differentiation with respect to the radial coordinate $r$.

The only components of $R_{\mu\nu}$ that do not vanish trivially are the diagonal components,
\begin{subequations}\label{eq:Ricci}
\begin{align}
    R_{tt} &=-\frac{A''}{2B}+\frac{A'}{4B}\para{\frac{A'}{A}+\frac{B'}{B}}-\frac{A'}{rB'},\label{eq:Rtt} \\
    R_{rr} &= \frac{A''}{2A}-\frac{A'}{4A}\para{\frac{A'}{A}+\frac{B'}{B}}-\frac{B'}{rB}, \label{eq:Rrr} \\
    R_{\theta\theta} &=\frac{1}{B}-1+\frac{r}{2B}\para{\frac{A'}{A}-\frac{B'}{B}},
    \label{eq:Rthth}
\end{align}
\end{subequations}
while $R_{\phi\phi} = R_{\theta\theta} \sin^2\theta$. By setting the above equations to zero, we arrive at the vacuum field equations. The equation for $R_{\phi\phi}$ gives the same information as the third one, so we will ignore it. By multiplying Eq.~\eqref{eq:Rtt} by $B/A$ and adding it to Eq.~\eqref{eq:Rrr}, we arrive at the following relation:
\begin{equation}
    A'B+AB'= \frac{d}{dr}(AB) = 0.
\end{equation}
We conclude that the product $AB = \alpha$ is constant. Substituting $B=\alpha/A$ in Eq.~\eqref{eq:Rthth}, we obtain
\begin{equation}
    A+rA'= \frac{d}{dr}(r A) = \alpha.
\end{equation}
Solving this equation, we get the following expressions for $A(r)$ and $B(r)$:
\begin{equation}
    \label{A(r) B(r)}
    A(r)=\alpha\para{1+\frac{\beta}{r}}, \qquad B(r)=\para{1+\frac{\beta}{r}}^{-1},
\end{equation}
where $\beta$ is another integration constant that results from solving the equation for $A(r)$.

The constants $\alpha$ and $\beta$ can be obtained from the weak-field limit:
\begin{equation}
    g_{tt} \simeq 1+\frac{2\Phi}{c^2},
\end{equation}
where $\Phi$ is the Newtonian gravitational potential. We thus have
\begin{equation}
    A(r) \xrightarrow[r \to \infty]{} 1 + \frac{2\Phi}{c^2} \qquad \text{(in weak-field limit)}
\end{equation}
For a spherically symmetric mass $M$, the Newtonian gravitational potential is $\Phi=-GM/r$ and we find that $\beta=-2GM/c^2$ and $\alpha=1$. The line element becomes:
\begin{equation}
    ds^2=\para{1-\frac{R_S}{r}}c^2 dt^2 - \para{1-\frac{R_S}{r}}^{-1}dr^2-r^2d\theta^2-r^2\sin^2\theta\hspace{3px}d\phi^2,
    \label{eq:ds2_Schwarzschild}
\end{equation}
representing the Schwarzschild solution, outside a spherically symmetric mass distribution, of the Einstein field equations. The quantity $\displaystyle R_S=\frac{2GM}{c^2}$ is known as the Schwarzschild radius.

In the following sections, we will consider bodies with their radius smaller than the Schwarzschild radius $R<R_S$. Such objects are called Schwarzschild black holes.

\section{Photon geodesics}\label{sec:geo}

In this section, we analyze photon geodesics on the Schwarzschild background spacetime. In Subsec.~\ref{sec:geo:general}, we derive the geodesic equations for a pointlike particle of arbitrary mass. Subsection~\ref{sec:geo:null} specializes these equations to the case of photon (null) geodesics. In Subsec.~\ref{sec:geo:stab}, we discuss the effective potential and the condition necessary for photons to be absorbed by the black hole. Finally, Subsec.~\ref{sec:geo:wf} discusses the weak-field limit, corresponding to large impact parameters, where the deflection angle can be easily obtained analytically.

\subsection{General geodesic equations} \label{sec:geo:general}

We calculated the Christoffel symbols \eqref{Christoffels_gen} for the general isotropic metric. Now knowing $A(r)$ and $B(r)$ from Eqs.~\eqref{A(r) B(r)}, we can write the Christoffel symbols for the Schwarzschild metric.
\begin{align}
\tensor{\Gamma}{^t_t_r} &= \frac{R_S}{2r^2}\para{1-\frac{R_S}{r}}^{-1},
& \tensor{\Gamma}{^r_t_t} &= \frac{c^2R_S}{2r^2}\para{1-\frac{R_S}{r}},
& \tensor{\Gamma}{^r_r_r} &= - \frac{R_S}{2r^2}\para{1-\frac{R_S}{r}}^{-1}, \notag \\
\tensor{\Gamma}{^r_\theta_\theta} &= -r\para{1-\frac{R_S}{r}},
& \tensor{\Gamma}{^r_\varphi_\varphi} &= -r\para{1-\frac{R_S}{r}}\sin^2\theta,
& \tensor{\Gamma}{^\theta_r_\theta} &= \frac{1}{r}, \notag \\
\label{Christoffels}
\tensor{\Gamma}{^\theta_\varphi_\varphi} &= -\sin\theta\cos\theta,
& \tensor{\Gamma}{^\varphi_r_\varphi} &= \frac{1}{r},
& \tensor{\Gamma}{^\varphi_\theta_\varphi} &= \cot\theta.
\end{align}

Now, we can write the geodesic equations
\begin{equation}
    \frac{d^2x^\mu}{d\lambda^2}+\tensor{\Gamma}{^\mu_\nu_\rho}\frac{dx^\nu}{d\lambda}\frac{dx^\rho}{d\lambda}=0,
\end{equation}
where $\lambda$ is an affine parameter. We obtain
\begin{subequations}\label{eq:geodesic_aux}
\begin{align}
    \frac{d}{d\lambda} \left[\para{1-\frac{R_S}{r}}\dot t\right] &= 0, \label{eq:geo_aux_t}\\
    \para{1-\frac{R_S}{r}}^{-1}\ddot r+\frac{R_S c^2}{2r^2}\dot t^2-\para{1-\frac{R_S}{r}}^{-2}\frac{R_S}{2r^2}\dot r^2-r\para{\dot\theta^2+\sin^2 \theta \hspace{3px} \dot\phi^2} &= 0, \\
    \ddot \theta + \frac{2}{r}\dot r \dot \theta -\sin\theta\cos\theta \hspace{3px}\dot\phi^2 &= 0, \label{eq:geodesic_aux_thth}\\
    \frac{d}{d\lambda} \left[r^2\sin^2\theta \hspace{3px} \dot\phi\right] &= 0, \label{eq:geo_aux_phi}
\end{align}
\end{subequations}
where the dot represents differentiation with respect to $\lambda$.

It can be seen that Eqs.~\eqref{eq:geo_aux_t} and \eqref{eq:geo_aux_phi} can be trivially integrated to give:
\begin{equation}
 \para{1-\frac{R_S}{r}}\dot t = k, \quad
 r^2\sin^2\theta \hspace{3px} \dot\phi = h.
 \label{eq:geo_aux_kh}
\end{equation}
The constants $k$ and $h$ can be interpreted physically in terms of the photon four-momentum $p_\mu$. This quantity can be introduced using the Lagrange formalism, starting from the Lagrangian $L=\frac{1}{2} g_{\mu\nu}\dot x^\mu \dot x^\nu$. The Euler-Lagrange equations,
\begin{equation}
\label{E-L eqn}
 \frac{d}{d\lambda} \frac{\partial L}{\partial \dot{x}^\mu} - \frac{\partial L}{\partial x^\mu} = 0,
\end{equation}
describe the motion of the photon and can be shown to coincide with Eqs.~\eqref{eq:geodesic_aux}.
In the Lagrangian formalism, the generalized momentum is defined by $p_\mu = \p L / \p \dot x^\mu$ \cite{bib:Taylor}, specifically:
\begin{equation}
\label{eq:pmu}
 p_t = \left(1 - \frac{R_s}{r}\right) c \dot{t}, \quad
 p_r = -\left(1 - \frac{R_s}{r}\right)^{-1} \dot{r}, \quad
 p_\theta = -r^2 \dot{\theta}, \quad
 p_\phi = -r^2 \sin^2 \theta \hspace{3px} \dot{\phi}.
\end{equation}

By virtue of Eq.~\eqref{E-L eqn}, the momentum components $p_\mu$ corresponding to cyclic coordinates $x^\mu$ (for which $\partial L / \partial x^\mu = 0$) are conserved. Comparing Eq.~\eqref{eq:geo_aux_kh} with Eq.~\eqref{eq:pmu}, we see that the constant $k$ and $h$ correspond to the particle energy, $p_t = E/c$, and $z$ component of the specific angular momentum, $p_\phi = -l^z$ \cite{bib:Hobson}:
\begin{equation}
 \label{physical interpretation of k and h}
 k = \frac{E}{c^2} = \frac{p_t}{c}, \quad
 h = l^z = - p_\phi.
\end{equation}

Due to the spherical symmetry of the geometry, we can work without loss of generality in the equatorial plane, $\theta= \pi / 2$. Then, Eqs.~\eqref{eq:geodesic_aux} become
\begin{subequations}
\begin{align}
    \label{first geo}
    \para{1-\frac{R_S}{r}}\dot t &= k, \\
    \para{1-\frac{R_s}{r}}^{-1}\ddot r+\frac{R_sc^2}{2r^2}\dot t^2-\para{1-\frac{R_s}{r}}^{-2}\frac{R_s}{2r^2}\dot r^2-r\dot\phi^2 &=0, \\
    \label{last geo}
    r^2 \dot\phi &= h,
\end{align}
\end{subequations}
while Eq.~\eqref{eq:geodesic_aux_thth} is automatically satisfied.
These are the general geodesic equations in the Schwarzschild geometry. Next, we will consider the geodesic equation for photons, which follow null curves.

\subsection{Null Geodesic equation} \label{sec:geo:null}

In the case of photons, we will work with Eqs.~\eqref{first geo}--\eqref{last geo}, where we will replace the complicated $r$ equation with the null curve condition $g^{\mu\nu} p_\mu p_\nu=0$:
\begin{subequations}\label{eq:null}
\begin{align}
    \label{null 1}
    \para{1-\frac{R_S}{r}}\dot t &=k, \\
    \label{null 2}
    \para{1-\frac{R_S}{r}} c^2 \dot t^2-\para{1-\frac{R_S}{r}}^{-1}\dot r^2 - r^2\dot\phi^2 &=0,\\
    \label{null 3}
    r^2\dot\phi &= h.
\end{align}
\end{subequations}
By substituting \eqref{null 1} and \eqref{null 3} into \eqref{null 2}, we arrive at the energy equation:
\begin{equation}
    \label{energy equation}
    \dot r^2+\frac{h^2}{r^2}\para{1-\frac{R_S}{r}}=c^2k^2.
\end{equation}

The trajectory equation for the photon can be obtained by considering $r = r(\phi)$ and performing the derivative with respect to $\lambda$ using the chain rule:
\begin{equation}
    \frac{dr}{d\lambda}=\frac{dr}{d\phi}\frac{d\phi}{d\lambda}=\frac{h}{r^2}\frac{dr}{d\phi}=-h\frac{d\para{1/r}}{d\phi}.
\end{equation}
Upon making the substitution $u=1/r$ and differentiating Eq.~\eqref{energy equation} with respect to $\phi$, we find
\begin{equation}
    \label{trajectory equation}
    \frac{d^2 u}{d\phi ^2}+ u =\frac{3R_S}{2} u^2,
\end{equation}
which is the trajectory equation for the photon, expressed in the form of Binet's equation encountered in the classical two-body problem \cite{bib:Taylor}.

\subsection{Stability of photon orbits}\label{sec:geo:stab}

\begin{figure}
    \centering
    \includegraphics[width=0.9\textwidth]{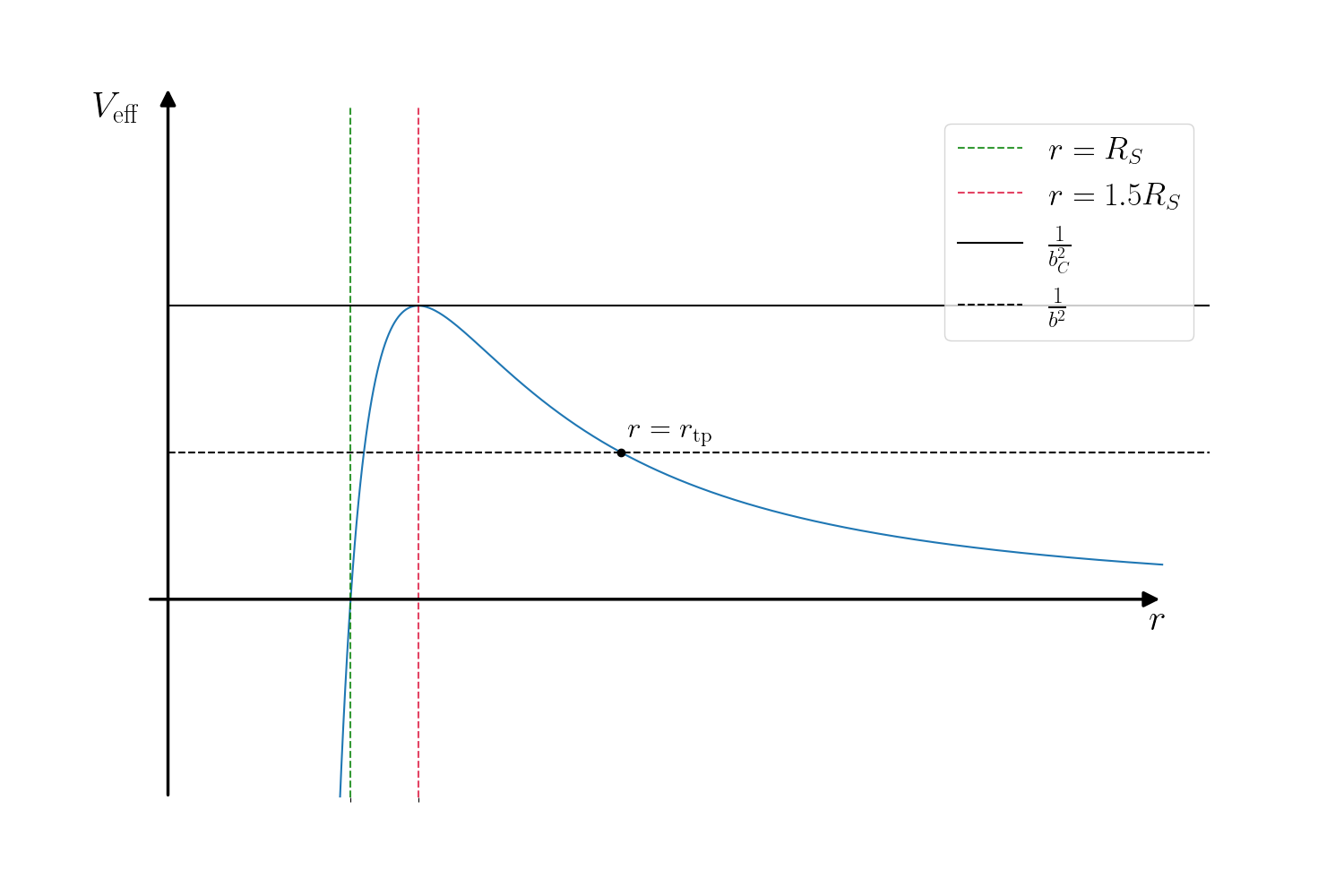}
    \caption{Effective potential shown in Eq.~\eqref{eq:Veff}. The crimson vertical dotted line represents the value of $r=1.5R_S$ at which the physical barrier (horizontal black line) is maximal. The green dotted line corresponds to the event horizon at $r=R_S$. The horizontal dotted line correspond to a photon coming at an impact parameter $b>b_C$ and the black circle at the intersection with the graph represents the turning point of the photon.
    }
    \label{fig:effective potential}
\end{figure}

The energy equation (\ref{energy equation}) can be written in the form
\begin{equation}
    \label{energy with effective potential}
    \frac{\dot r^2}{h^2}+V_{\text{eff}}(r)=\frac{1}{b^2},
\end{equation}
where we defined the impact parameter $b=h/(ck)$ and the effective potential
\begin{equation}
    V_{\text{eff}}(r)=\frac{1}{r^2}\para{1-\frac{R_S}{r}}.
    \label{eq:Veff}
\end{equation}
It can be seen that the effective potential has a maximum at $r=1.5R_S$, where it takes the value $V_{\rm eff}^{\rm max} = \displaystyle \para{\frac{3\sqrt{3}}{2}R_S}^{-2}$. Thus, the circular orbit at $r=1.5R_S$ is unstable. As can be seen from the shape of the effective potential, shown with respect to $r$ in Fig.~\ref{fig:effective potential}, there are no stable circular orbits.

Writing now Eq.~\eqref{energy with effective potential} in the following form,
\begin{equation}
    \frac{\dot r^2}{h^2}=\frac{1}{b^2}-V_{\text{eff}}(r), \label{eq:drsq_aux}
\end{equation}
we can interpret the left-hand side as the kinetic energy of the photon, which is always a positive real number. If there exists a region where $V_{\text{eff}}>1/b^2$, then the left-hand side of Eq.~\eqref{eq:drsq_aux} would be negative, which is physically impossible. The term $1/b^2$ acts as the total energy of the photon.

If there are values of the radius for which $V_{\rm eff}(r) > 1/b^2$, the trajectory must have a turning point $r_{\rm tp}$ where $\dot r=0$ (an example illustrated in Fig.~\ref{fig:effective potential}) and the photon will escape to infinity because it cannot cross the physical barrier $V_{\text{eff}}=1/b^2$. The minimum impact parameter at which this can happen is
\begin{equation}
 b_C = \frac{3\sqrt{3}}{2} R_S \simeq 2.6 R_S,
 \label{eq:bC}
\end{equation}
since $V_{\text{eff}}^{\text{max}} = 1/b_C^2$. If the impact parameter is smaller than this critical value, $b<b_C$, then the photon has enough energy to cross the physical barrier $V_{\text{eff}}^{\text{max}}=1/b_C^2$ and its trajectory will not have a turning point, therefore the photon will be absorbed by the black hole.

\subsection{Weak-field approximation}\label{sec:geo:wf}

By scaling our trajectory equation with the impact parameter $b$ such that $\tilde{u} = u b$, the trajectory equation becomes dimensionless:
\begin{equation}
    \label{scaled trajectory}
    \frac{d^2 \tilde u}{d\phi ^2}+ \tilde u =\frac{3R_S}{2b} \tilde u^2.
\end{equation}

In the absence of matter, $R_S = 0$, the right-hand side of the trajectory equation (\ref{scaled trajectory}) vanishes:
\begin{equation}
    \frac{d^2 \tilde u_0}{d\phi ^2}+ \tilde u_0=0.
\end{equation}
In this case, the photon will travel on a straight line and we can write the solution using the impact parameter $b$ satisfying our initial conditions:
\begin{equation}
 \left. u_0 \right\rvert_{\phi = 0} = \left.\tilde{u}_0\right\rvert_{\phi = 0} = 0, \quad
 \left.\frac{du_0}{d\phi}\right\rvert_{\phi = 0} = \frac{1}{b}\Longleftrightarrow \left. \frac{d\tilde u_0}{d\phi}\right\rvert_{\phi=0}=1,
 \label{eq:ic}
\end{equation}
which give:
\begin{equation}
    \tilde u_0 = \sin \phi \Longleftrightarrow u_0=\frac{\sin\phi}{b}.
\end{equation}

Next, we consider the weak-field approximation $g_{\mu\nu}=\eta_{\mu\nu}+h_{\mu\nu}$, where $\eta_{\mu\nu}$ is the flat Minkowski metric and $|h_{\mu\nu}|\ll 1$. In our trajectory equation (\ref{scaled trajectory}), this condition imposes that the coefficient of the $\tilde u^2$ term is small, such that
\begin{equation}
    \frac{d^2 \tilde u}{d\phi^2}+\tilde u=\varepsilon \tilde u^2, \qquad \text{with} \quad\varepsilon=\frac{3R_S}{2b}.
\end{equation}
We now consider a perturbation to the flat-space solution of the form
\begin{equation}
    \tilde u = \tilde u_0 + \varepsilon \tilde u_1 + O(\varepsilon^2).
\end{equation}
Substituting the above into the trajectory equation \eqref{scaled trajectory}, we arrive at
\begin{equation}
    \frac{d^2 \tilde u_0}{d\phi^2}+\varepsilon  \frac{d^2 \tilde u_1}{d\phi^2}+\tilde u_0 +\varepsilon \tilde u_1=\varepsilon\tilde u_0^2 + O(\varepsilon^2).
\end{equation}
Using the fact that $\tilde u_0$ is the flat space solution, we are left with the following ODE:
\begin{equation}
    \frac{d^2 \tilde u_1}{d\phi^2} + \tilde u_1 = \tilde u_0^2=\sin^2\phi.
\end{equation}
Using the trigonometric identity $\sin^2 \phi=(1-\cos2\phi)/2$, we can guess the particular solution for $\tilde u_1$ of the form $\tilde{u}^{\rm p}_1=A\cos 2\phi +B\sin 2\phi + C$. Substituting in the ODE, we find that $\tilde{u}^{\rm p}_1$ is
\begin{equation}
    \tilde{u}^{\rm p}_1=\frac{1}{2}+\frac{1}{6}\cos 2\phi.
\end{equation}
The homogeneous solution, $\tilde u_1^{\rm h} = \alpha \sin \phi + \beta \cos\phi$, can be used to impose the initial conditions in Eq.~\eqref{eq:ic}, leading to
\begin{equation}
 \alpha = 0, \qquad \beta = -\frac{2\varepsilon}{3}.
\end{equation}
Therefore, the full solution of $\tilde u$ in the weak-field approximation is
\begin{equation}
    \label{perturbed solution}
    \tilde u= \sin \phi + \varepsilon \left(\frac{1}{2} - \frac{2}{3} \cos\phi + \frac{1}{6} \cos 2\phi\right).
\end{equation}

\begin{figure}
    \centering

\begin{tikzpicture}[>=latex, font=\normalsize]

  \fill[black] (0,0) circle (0.25);

  \draw[dashed, thick] (-5.5, 0) -- (5.5, 0);

  \draw[dashed, thick] (-5.5, 1.8) -- (5.5, 1.8);

  \draw[<->, thick] (4.5, 0) -- (4.5, 1.8) node[midway, left=2pt] {$b$};

  \draw[thick] plot[smooth, domain=-5.5:5.5]
    (\x, {1.8 - 0.03*max(0, -\x + 1.0)^2});

  \draw[->, thick] (-5.0, 1.8) arc (0:13:-4.6) node[midway, right=2pt] {$\delta \phi$};

\end{tikzpicture}

\caption{Geometry of the trajectories (not to scale). The upper dotted line represents the unperturbed path and the continuous line represents the actual trajectory (greatly exaggerated) around the body. The quantity $\delta \phi = \phi_\infty - \pi$ represents the deflection angle.}
\label{fig:deflection angle}

\end{figure}
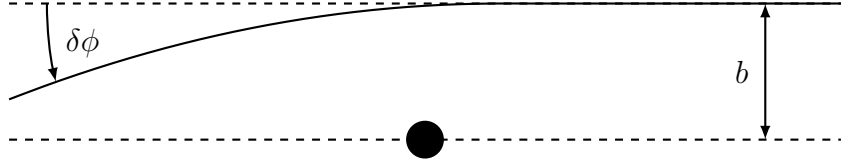

In flat space, if we consider a photon coming from the right ($\phi=0$) at an impact parameter $b$ parallel to the observation axis, the photon will continue its journey unperturbed and escape at $\phi=\pi$. We want to see the angle at which the photon escapes in the weak-field limit, $\phi_\infty = \pi + \delta \phi$, so we will take the limit $ \tilde u\to  0$ ($r \to \infty$) and use the small angle approximation for the trigonometric functions
\begin{equation}
    -\delta \phi +\varepsilon\para{\frac{1}{2} + \frac{2}{3} +\frac{1}{6}} = 0.
\end{equation}
The deflection angle $\delta \phi$ represents the angle between the escaping path of the unperturbed photon and the bent path, as shown in the sketch in Fig.~\ref{fig:deflection angle}. In the weak-field approximation, this is given by
\begin{equation}
    \label{weak field angle}
    \delta\phi = \frac{4}{3}\varepsilon = \frac{2R_S}{b} = \frac{4GM}{c^2 b},
\end{equation}
being proportional to the perturbative parameter $\varepsilon$ (or $R_S/b$).

\section{Numerical results}\label{sec:num}

The non-linearity of Eq.~\eqref{trajectory equation} greatly amplifies the difficulty of writing an analytic solution. Perturbation theory provides useful approximations in different weak field scenarios (like photons around the sun) but is not of much help in more extreme cases like in the presence of a black hole.

An analytical solution involving elliptic functions exists \cite{bib:Darwin,bib:Bozza2010}:
\begin{equation}
 \delta \phi_{\rm exact} = -\pi + 4\sqrt{\frac{r_m}{s}}F(\varphi,m),
 \label{eq:phi_exact}
\end{equation}
where $F(\varphi,m) = \int_0^\varphi d\theta / \sqrt{1 - m^2 \sin^2\theta}$ \cite{DLMF} is the incomplete elliptic integral of the first kind. The constants are defined by the following relations:
\begin{align}
    s &=\sqrt{(r_m-2R_s)(r_m+3R_s)}, \notag \\
    m &=\frac{s-r_m+3R_s}{2s}, \notag \\
    \varphi &=\arcsin{\sqrt{\frac{2s}{3r_m - 3R_S + s}}}. \notag
\end{align}
In the above, $r_m$ represents the radius of the closest approach, obtained by setting $\dot{r} = 0$ in Eq.~\eqref{energy equation}:
\begin{equation}
 \left(\frac{R_S}{b}\right)^2 \left(\frac{r}{R_S}\right)^3 - \frac{r}{R_S} + 1 = 0.
\end{equation}

Equation~\eqref{eq:phi_exact} is cumbersome to use in order to determine the photon trajectory. Instead, for analyzing and visualizing trajectories, we turn to numerically solving equation \eqref{trajectory equation} near a Schwarzschild black hole using an in-house solver written in Python, employing the fourth-order Runge-Kutta integration scheme \cite{bib:Butcher}, described below.

We first write the second-order differential equation~\eqref{scaled trajectory} as a coupled set of first-order differential equations:
\begin{equation}
 \frac{dU}{d\phi} = L(U, \phi), \quad
 U = \begin{pmatrix}
 \tilde{u} \\ \tilde{u}'
 \end{pmatrix}, \quad
 L = \begin{pmatrix}
 \tilde{u}' \\
 \dfrac{3R_S}{2b} \tilde{u}^2 - \tilde{u}
 \end{pmatrix}.
\end{equation}
The integration routine is started from $\tilde{u}_0 = 0$ and $\tilde{u}'_0 = 1$ at $\phi_0 = 0$ and constructs the solution $U_n$ at $\phi_n$ via the fourth-order Runge-Kutta algorithm. Considering the solution $U_n$ known, the solution at $\phi_{n+1} = \phi_n + \Delta \phi$ is obtained via
\begin{equation}
\label{rk4}
 U_{n+1} = U_n + \frac{\Delta \phi}{6} (k_1 + 2k_2 + 2k_3 + k_4),
\end{equation}
where the intermediate steps are given as
\begin{align}
\label{rk4 steps}
 k_1 &= L(U_n, \phi_n), &
 k_2 &= L\left(U_n + \frac{\Delta \phi}{2} k_1, \phi_n + \frac{\Delta \phi}{2}\right), \nonumber\\
 k_3 &= L\left(U_n + \frac{\Delta \phi}{2} k_2, \phi_n + \frac{\Delta \phi}{2}\right), &
 k_4 &= L\left(U_n + \Delta \phi k_3, \phi_n + \Delta \phi\right).
\end{align}
We perform a number of $N$ steps, until the $N+1$'th step yields $\tilde{u}_{N+1} < 0$. This signals that the angle $\phi_{N+1} = \phi_N + \Delta \phi$ exceeds the escape angle $\phi_\infty$.

In order to obtain the escape angle, we employ the following algorithm. We consider performing RK4 iterations with decreasing angular steps, $\Delta \phi^{(j)} = \Delta \phi / 2^j$, with $1 \le j \le J$. For each value of $j$, we apply Eq.~\eqref{rk4} with the angular step $\Delta \phi^{(j)}$, as long as the resulting $\tilde{u}_{N+1}^{(j)}$ remains positive. If $\tilde{u}_{N+1}^{(j)}$ is larger than a threshold (we employed $10^{-15}$), then the value of $j$ is incremented by one unit and the process is repeated. We consider $\tilde{u}_{N + 1} = \tilde{u}_{N+1}^{(J)}$ and the subsequent escape angle $\phi_\infty = \phi_{N+1}^{(J)}$.

\begin{figure}[h]
    \centering
    \begin{tabular}{cc}
 \includegraphics[width=.45\textwidth]{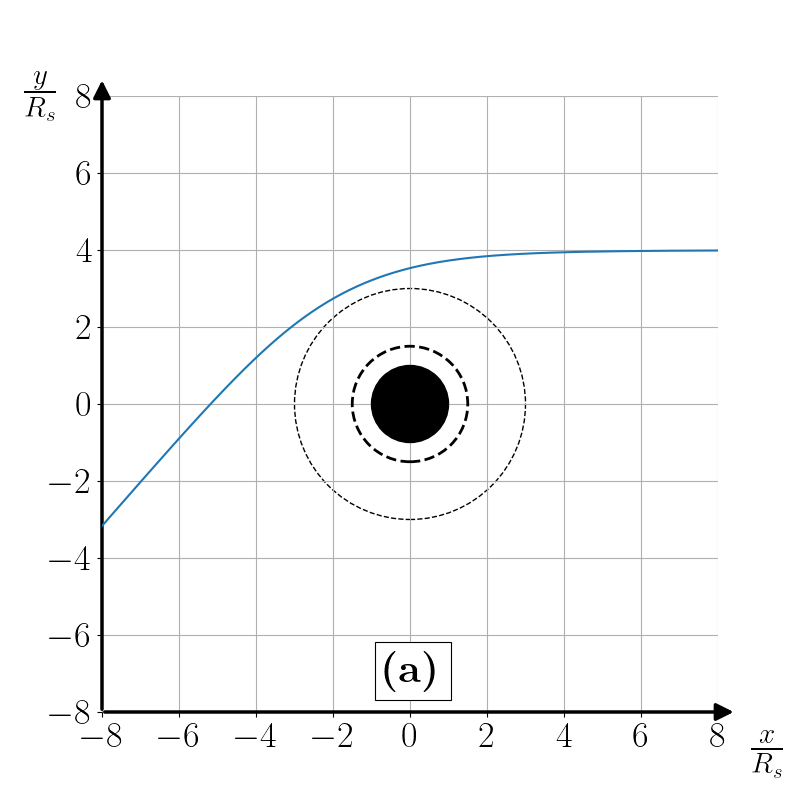} &
 \includegraphics[width=.45\textwidth]{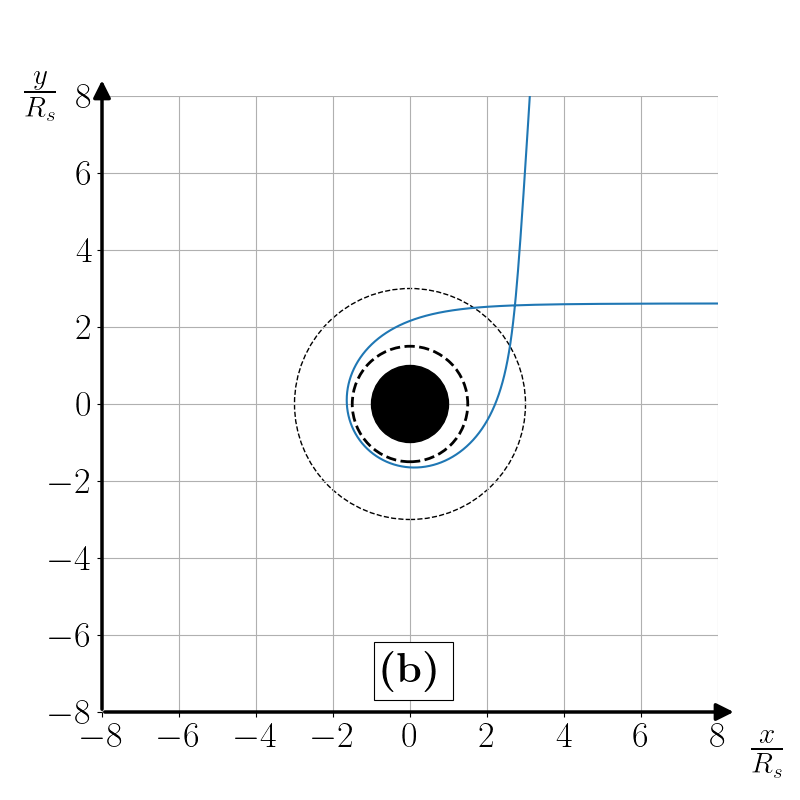} \\
 \includegraphics[width=.45\textwidth]{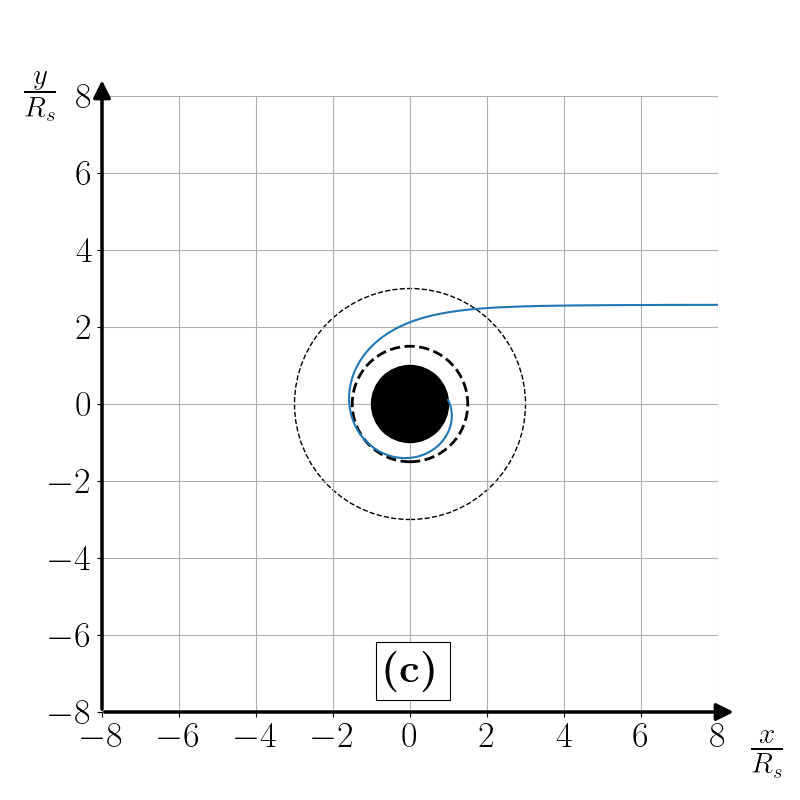} &
 \includegraphics[width=.45\textwidth]{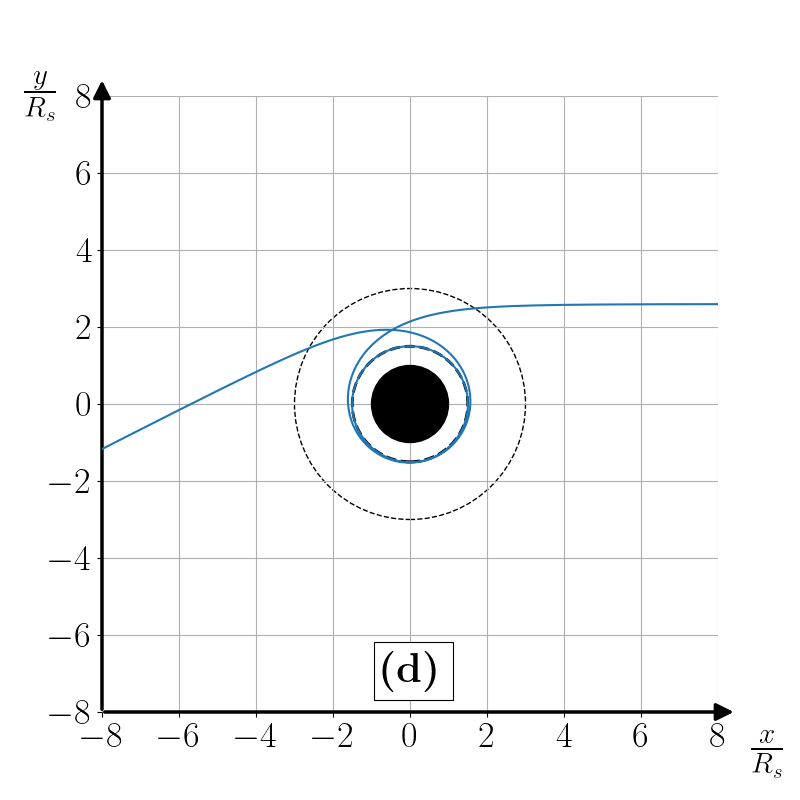}
 \end{tabular}
    \caption{
    Sample photon trajectories, shown with blue lines, for impact parameters $b=4R_S$ (a), $2.615 R_S$ (b), $2.58 R_S$ (c) and $2.59808 R_S$ (d). The black disk corresponds to the black hole, extending up to $r = R_S$. The outer dotted circle represents the ISCO at $r = 3R_S$ and the inner dotted circle represents the photon sphere at $r = 1.5R_S$.
    \label{fig:Trajectories}
    }
\end{figure}

In Fig~\ref{fig:Trajectories}, we show sample trajectories for which the impact parameters are chosen to showcase smaller (a), or higher (b) bending of the photon path, a photon falling inside the event horizon (c) and a photon doing 2 loops around the black hole (d).

For a discrete numerical solution obtained with angular step $\Delta \phi$, the global error is
\begin{equation}
    \delta \phi(\Delta \phi) - \delta \phi_{\text{analytic}} \simeq a\Delta \phi^{\gamma},
\end{equation}
where $\gamma=4$ in the case of an RK4 scheme. Upon taking the base-10 logarithm of both sides of the above equation, we have
\begin{equation}
    \label{log error}
    \lg\para{\delta \phi - \delta \phi_{\text{analytic}}} \simeq \lg a + \gamma \lg \Delta \phi.
\end{equation}

\begin{figure}[h]
\begin{tabular}{cc}
 \includegraphics[width=.48\textwidth]{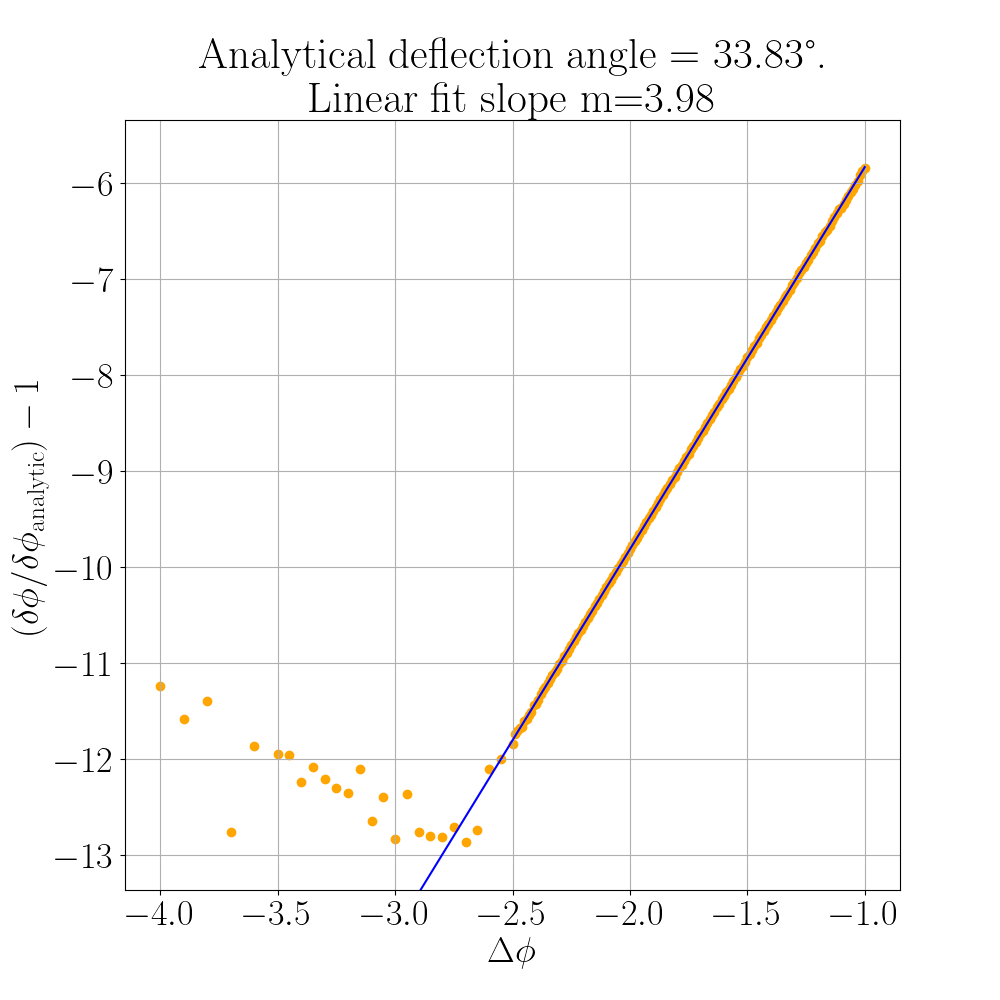} &
 \includegraphics[width=.48\textwidth]{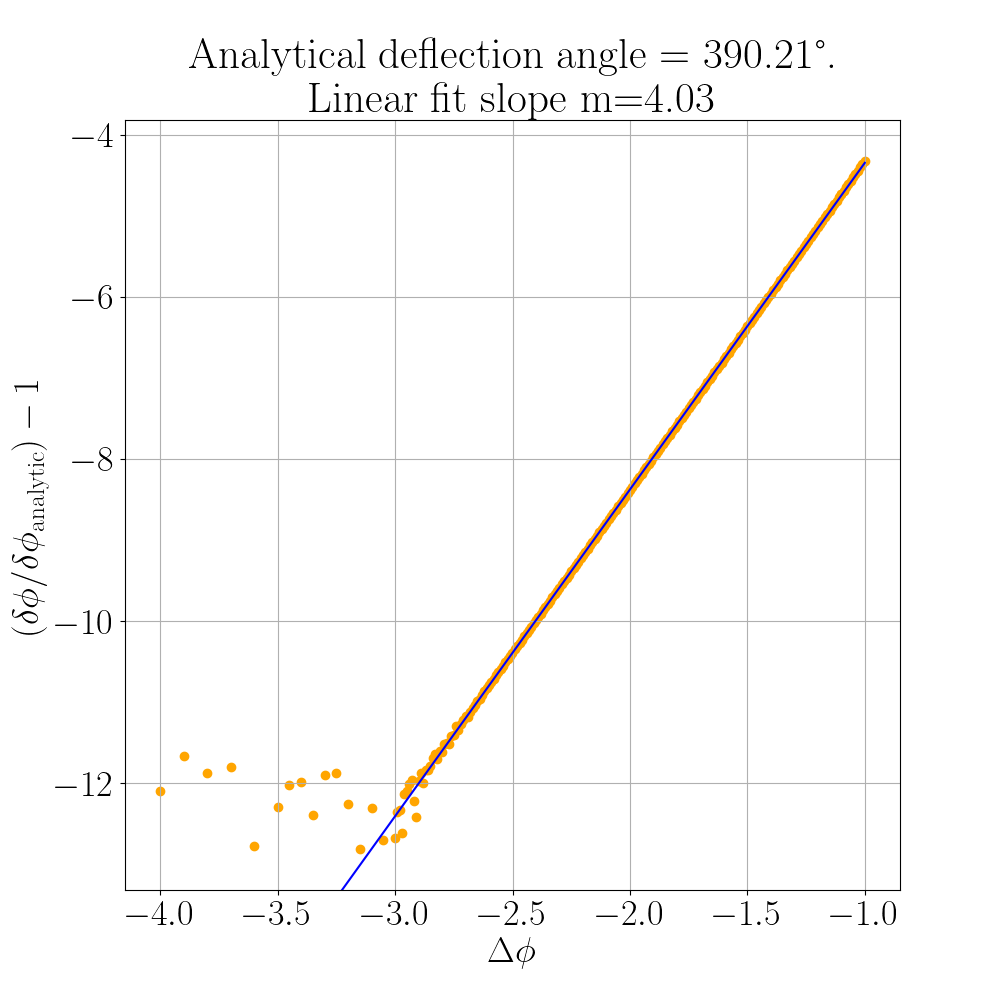} \vspace{-45pt} \\
 (a) & (b) \vspace{20pt} \\
 \includegraphics[width=.48\textwidth]{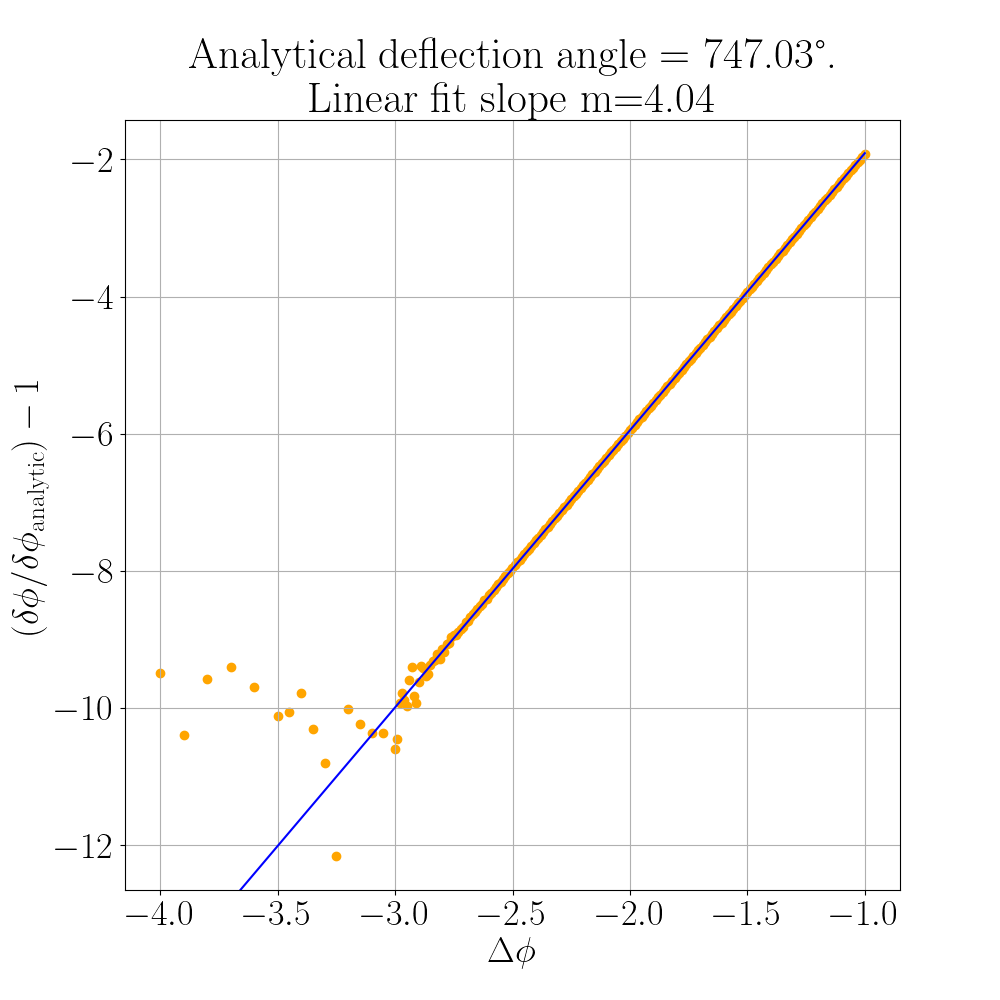} &
 \includegraphics[width=.48\textwidth]{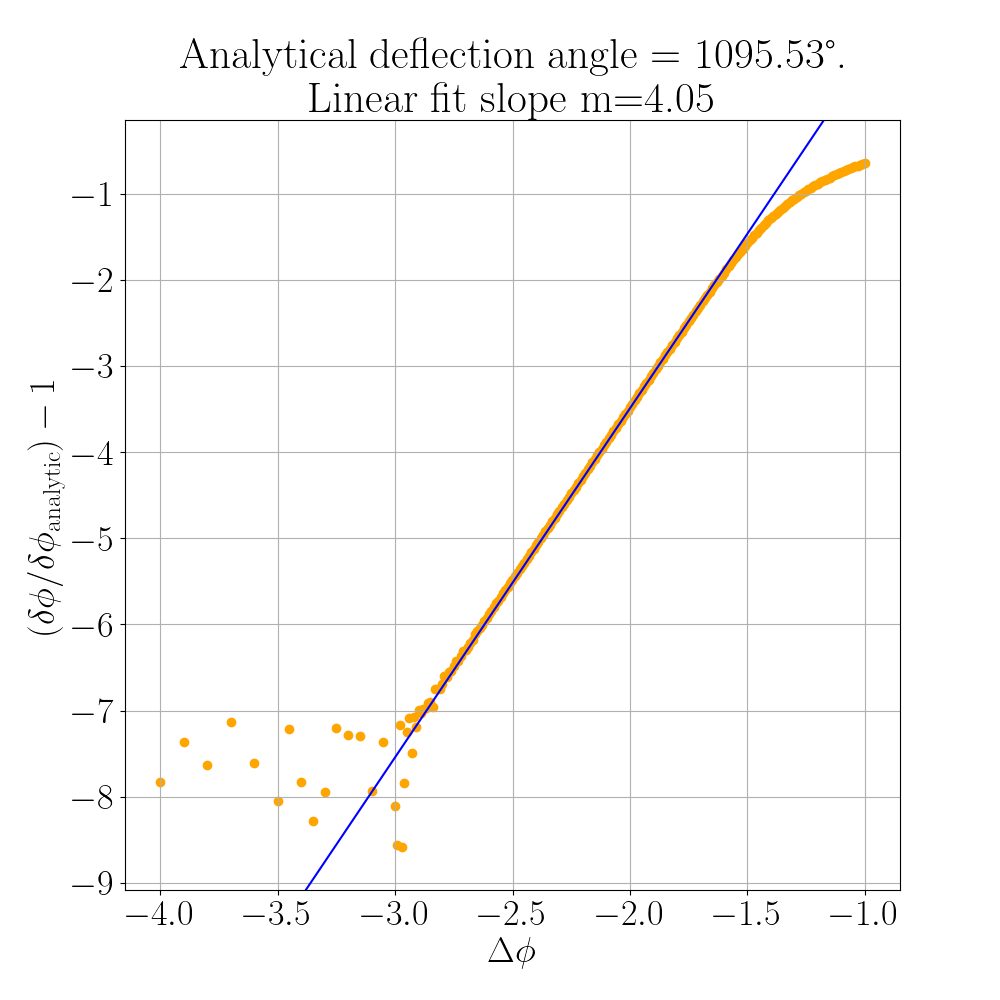} \vspace{-45pt} \\
 (c) & (d) \vspace{20pt}
\end{tabular}
 \caption{Log-log plot of the convergence of the numerically-determined deflection angle using the RK4 scheme to the analytical result in Eq.~\eqref{eq:phi_exact}, represented with respect to the angular step $\Delta \phi$. Both axes are on base-10 logarithmic scale. The impact parameters for which the analysis was carried are: (a) $5R_S$, (b) $2.6R_S$ corresponding to a photon doing a loop, (c) $2.598080 R_S$ corresponding to a photon doing 2 loops, (d) $2.59807622 R_S$ corresponding to a photon doing 3 loops.
 \label{fig:Convergence plots}
 }
\end{figure}

An analysis of the convergence of the solution to the analytical solution \eqref{eq:phi_exact} confirms the fourth-order convergence (slope $\gamma=4$ in \eqref{log error}) of the computed deflection angle to the exact solution when the angular step $\delta \phi$ decreases. The numerical limit, at which floating point number errors become dominant, is quickly reached, at angular steps around $\Delta \phi=10^{-3}$. The accuracy of the simulated trajectory slowly decreases as the photon goes around the black hole multiple times. Simulations were carried out for various values of the impact parameter $b$, which are shown in Fig.~\ref{fig:Convergence plots}.

A linear fit $y=m\cdot x + x_0$ through the points which are not affected by accumulated floating number errors yields, as seen in Fig~\ref{fig:Convergence plots}, a slope around $\gamma\simeq4$, as expected for a fourth order scheme.

\begin{figure}[h]
    \centering
    \includegraphics[width=1\textwidth]{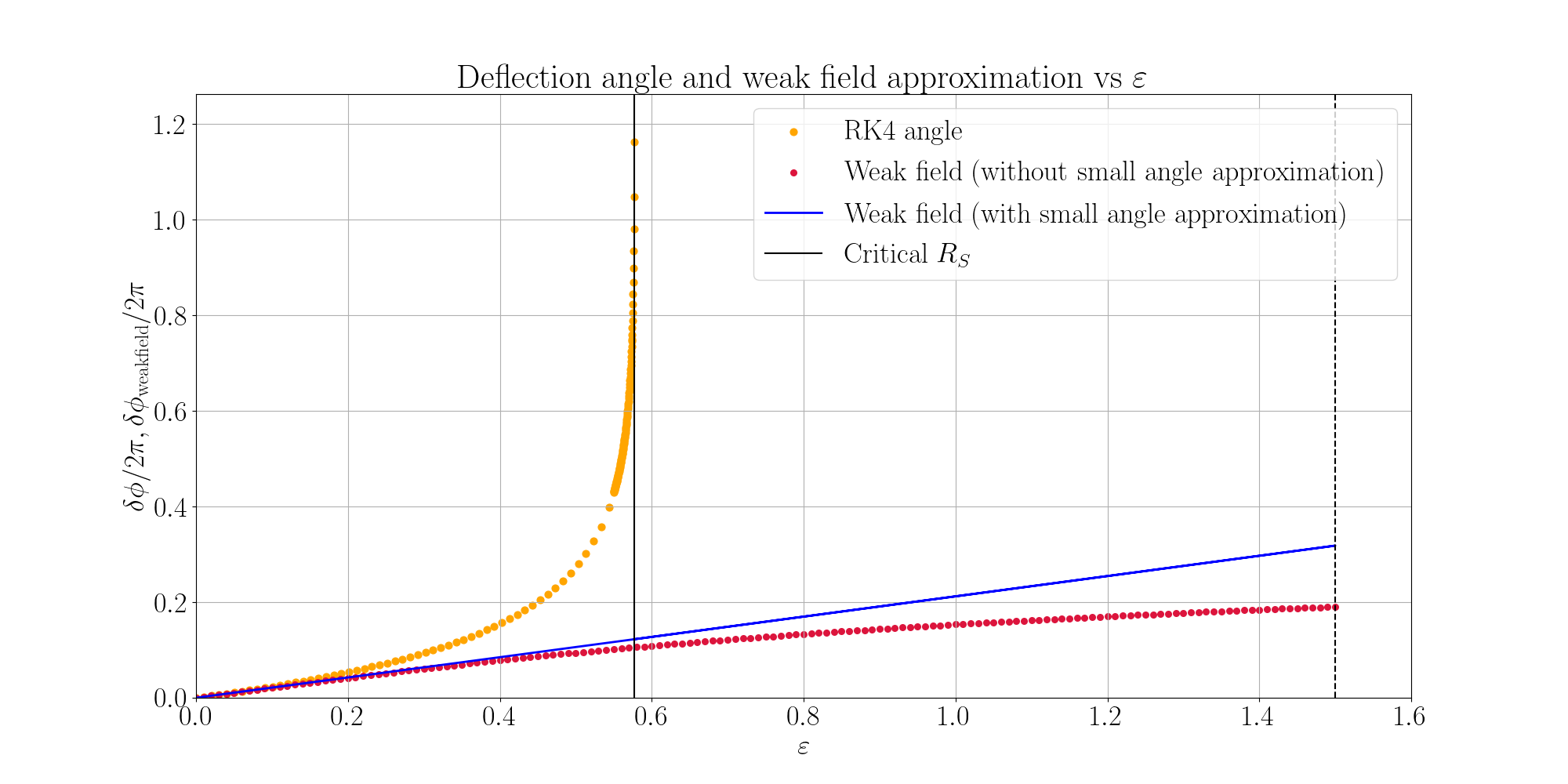}
    \caption{Simulated deflection angle $\delta \phi$ (the orange dots), the weak field approximation involving the small angle approximation in Eq.~\eqref{weak field angle} given by the blue line and the weak field formula without involving the small angle approximation ($\tilde u = 0$ in \eqref{perturbed solution}) represented by crimson dots, as functions of the parameter $\varepsilon = 3R_S / 2b$. The angles on the y axis are in multiples of $2\pi$. Angles exceeding $2\pi$ correspond to deflection of photons that loop around the black hole.
    }
    \label{fig:weak field and simulated deflection}
\end{figure}

Considering the scaled trajectory equation \eqref{scaled trajectory}, we can check the validity of the weak-field approximation \eqref{weak field angle}. From Fig.~\ref{fig:weak field and simulated deflection}, we can see that the weak-field formula \eqref{weak field angle} is a good approximation only for very small deflections. The weak field approximation, without expanding the trigonometric functions as power series, by setting $\tilde u = 0$ in equation \eqref{perturbed solution}, is a worse approximation in general compared to \eqref{weak field angle}.

We also employed a binary search to find the impact parameter at which the photons start to be absorbed by the black hole. This is represented as a vertical solid black line in Fig.~\ref{fig:weak field and simulated deflection}. The value for $b_C$ found numerically agrees with the theoretical value in Eq.~\eqref{eq:bC} from the stability analysis of photon orbits in Subsec.~\ref{sec:geo:stab} within numerical precision (up to 14 decimals).

\section{Black hole image}\label{sec:bh_image}

With the technological advance of recent years, like the Event Horizon Telescope \cite{bib:Event Horizon Telescope}, directly capturing images of black holes became a real possibility \cite{bib:EHT2019,bib:EHT2022SgrA}. Of course, the black hole itself is invisible to external observers, however the accretion disk around it casts a unique image. Testing different models and theoretical ideas with these images is crucial to both interpret the astrophysical images and to verify the accuracy of our theoretical models.

Due to gravitational lensing, the trajectory of photons emitted from behind the black hole are bent and can reach the observer, giving rise to an apparent ring above the black hole event horizon. As already seen in the previous section, some photons arriving under impact parameters below a threshold end up falling into the black hole. Therefore, the black hole casts a shadow (a dark disk) of radius larger than that of the black hole itself, $R_{\rm shadow} \simeq \frac{3}{2} R_S \sqrt{3} \simeq 2.6 R_S$.

In this section, we use our numerical solver for photon trajectories to simulate images, as seen by a distant observer, of accretion disks around black holes. Subsection~\ref{sec:bh_image:redshift} discusses the aspects of the gravitational redshift that the far observer perceives. Subsection~\ref{sec:bh_image:disk} presents the model of the accretion disk we adopted and the general geometry of the problem. Subsection~\ref{sec:bh_image:head-on} touches over the particular case of viewing the black hole accretion disk head-on (under an inclination angle $i=0$) while Subsection~\ref{sec:bh_image:incline} goes over the more general case of an inclination angle $i \neq 0$. Subsection~\ref{sec:bh_image:rot} discusses the situation in which the matter of the accretion disk follows the Keplerian circular orbit and the additional redshift that arises due to the rotation of the disk.

\subsection{Gravitational redshift}\label{sec:bh_image:redshift}

Photons travelling near a black hole will experience a gravitational redshift (or blueshift) that is perceived by the observer. The redshift of a photon in stationary spacetime ($\p_0g_{\mu\nu}=0$) is given by \cite{bib:Hobson}
\begin{equation}
    \frac{\nu_R}{\nu_E}=\rpara{\frac{g_{00}(x_E)}{g_{00}(x_R)}}^{1/2},
\end{equation}
where $\nu_R$ and $\nu_E$ are the received and emitted frequencies, respectively, while $x_R$ and $x_E$ represent the coordinates of the receiver and of the point of emission, respectively.
For the Schwarzschild metric given in Eq.~\eqref{eq:ds2_Schwarzschild}, we have
\begin{equation}
 \label{redshift}
 \frac{\nu_R}{\nu_E}=\rpara{\frac{1-R_S/r_E}{1-R_S/r_R}}^{1/2} \xrightarrow[r_R \to \infty]{}
 \para{1-\frac{R_S}{r_E}}^{1/2},
\end{equation}
where we took into account that the receiver is far away from the black hole.

There is an additional cosmological redshift, due to the expansion of spacetime, $\nu_{\rm obs} = \nu_R(1 + z)$, where $z$ represents the redshift factor associated with the emission region. For example, for the M87 black hole, $z \simeq H_0 D / c \simeq 0.004$ for the distance $D = 16.8$ Mpc \cite{bib:EHT2019} and the Hubble constant $H_0 \simeq 70\ {\rm km}/{\rm s}/{\rm Mpc}$ \cite{Farooq:2012ev,Cai:2026swf}, which gives a negligible redshift. Since this redshift applies unitarily to all photons, we will not take it into account explicitly in what follows.

In the following section, we consider impact parameters of the test photons that are expressed as multiples of the Schwarzschild radius $R_S$. Thus, the image can represent a black hole of any mass.

\subsection{Accretion disk model and geometry} \label{sec:bh_image:disk}

In creating the images, we employed the $\alpha$-disk model proposed by Shakura and Sunyaev \cite{Shakura:1972te,bib:Accretion}. The model assumes the disk to be in thermal equilibrium.
The temperature distribution on the disk according to this model is \cite{bib:Accretion}
\begin{equation}
    \label{temperature distribution}
    T_C=1.4 \cdot 10^4 \alpha^{-1/5} \dot M_{16}^{1/3} m^{1/4}_1 R_{10}^{3/4} f^{6/5} K,
\end{equation}
where $\alpha$ is a free parameter, $\dot M_{16}$ is the accretion rate in units of $10^{16} \text{g}\hspace{3px} \text{s}^{-1}$, $m_1$ is the mass of the central object in units of solar masses $M_\odot$, $R_{10}$ is the radius at a point on the disk in units of $10^{10}$ cm and $f$ is given by
\begin{equation}
    f=\rpara{1-\para{\frac{R_\star}{R}}^{1/2}}^{1/4}.
\end{equation}
Here, $R_\star$ is the radius where angular momentum stops being transported inward. In our case $R_\star = 3R_S$.
The disk was assumed to radiate according to the blackbody radiation spectrum. Both the case in which the disk is not rotating and the one in which it has angular momentum are considered.

\begin{figure}[h]
    \centering
\begin{tabular}{cc}
\begin{tikzpicture}[scale=1.2]

            \draw[->, thick] (-3, 0) -- (3.2, 0) node[right] {$z$};
            \draw[->, thick] (0, -2.5) -- (0, 3.125) node[above] {$y$};

            \fill[black] (0, 0) circle (0.35);

            \draw[orange, ultra thick] (-1.5, -0.5) -- (1.5, 0.5);

            \draw[blue, ultra thick, ->] (0.9,0.3) -- (0.6,1.2);
            \draw[blue, ultra thick, ->] (-0.9,-0.3) -- (-1.2,0.6);
            \draw[blue] (-1.5, 0.6) node {$\vec{n}$};

            \draw[dashed, thick] (-2.8, -2.5) .. controls (-1, 0) and (-0.6, 1.2) .. (2.8, 1.2);

            \draw[->, thick] (1.3, 0) arc (0:12:2.0) node[midway, right=2pt, yshift=-2pt] {$i$};

        \end{tikzpicture}
        &
        \begin{tikzpicture}[scale=1.5]

            \coordinate (O) at (0,0);
            \coordinate (P) at ({90-30}:2.2);
            \coordinate (Py) at (0, {2.2*cos(30)});
            \coordinate (Pz) at ({2.2*sin(30)}, 0);

            \draw[->, thick] (-2, 0) -- (2.5, 0) node[right] {$x$};
            \draw[->, thick] (0, -2) -- (0, 2.5) node[above right] {$y$};

            \draw[dashed, thick] (P) -- (Py);
            \draw[dashed, thick] (P) -- (Pz);

            \draw[line width=1.2pt] (O) -- (P);

            \draw[dashed,->, thick] (0, 0) -- ({90-30}: 3) node[right]{$y'$};

            \draw[dashed,->, thick] (0, 0) -- ({-30}: 2.7) node[right]{$x'$};

            \draw[->, thick] (0,1) arc (90:50:0.7) node[midway, right=-4pt, yshift=5pt] {$\varphi$};

            \draw[->, thick] (2,0) arc (0:-30:1.9) node[midway, right=2pt] {$\varphi$};

            \node at (45:1.4) {$b$};

            \fill (O) circle (0.28);

        \end{tikzpicture}
    \end{tabular}
    \caption{(Left) Black hole side view. The observer is at $z \to \infty$ and the accretion disk is seen as the line shown in orange, with its normal, shown in blue, contained in the $yOz$ plane. The dotted line illustrates the photon trajectory. (Right) Observer view (along the $z$ axis) of the black hole. The photon trajectory is contained in the $y'Oz$ plane.}
    \label{fig: Trajectory and disk geometry}

\end{figure}
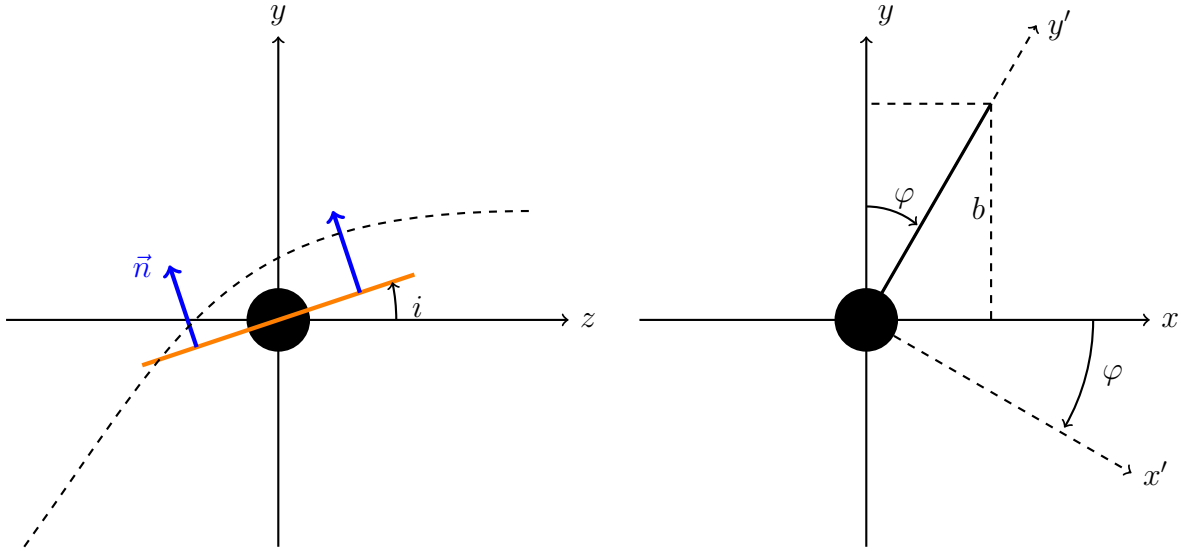

In Fig.~\ref{fig: Trajectory and disk geometry}, a sketch of the general geometry of the disk (for arbitrary accretion disk inclination angle $i$) and the trajectory is presented.
If a photon comes at a nonvanishing value of the $x$ coordinate, due to the spherical symmetry of the problem, we can perform a rotation of the system around the $z$ axis so that the trajectory is fully embedded in the plane $x'=0$ (the prime denotes the rotated coordinates). The new coordinates can be written using the rotation matrix $R(\varphi)$:
\begin{equation}
    \begin{pmatrix}
        x'\\y'
    \end{pmatrix}=
    \begin{pmatrix}
        \cos\varphi & -\sin\varphi \\
        \sin \varphi& \cos \varphi
    \end{pmatrix}
    \begin{pmatrix}
        x\\y
    \end{pmatrix}.
    \label{eq:varphi_def}
\end{equation}

The old coordinates can be calculated back using the inverse rotation matrix $R^{-1}(\varphi)$:
\begin{equation}
    \label{inverse rotation}
    \begin{pmatrix}
        x\\y
    \end{pmatrix}=
    \begin{pmatrix}
        \cos\varphi & \sin\varphi \\
        -\sin \varphi & \cos \varphi
    \end{pmatrix}
    \begin{pmatrix}
        x'\\y'
    \end{pmatrix}.
\end{equation}

We can describe the plane of the accretion disk via $\vec{n}\cdot\vec{r}=d$, where, by imposing the disk plane to contain the origin, we obtain $d=0$. The disk's normal is, according to Fig.~\ref{fig: Trajectory and disk geometry} (left panel),
\begin{equation}
    \vec{n}= \cos{i}\hspace{3px}\vec{j} - \sin i \hspace{3px}\vec{k},
\end{equation}
where $i$ represents the inclination angle of the disk with respect to the observer's line of sight.
We can therefore write the equation for the accretion disk's plane as
\begin{equation}
    \vec{n}\cdot\vec{r}=0 \Longrightarrow y\cos i - z\sin i=0.
\end{equation}
Using the inverse rotation \eqref{inverse rotation}, we arrive at
\begin{equation}
    -x'\cos i \sin\varphi+y'\cos i\cos\varphi-z'\sin i=0.
\end{equation}

The photon moves entirely in the $x'=0$ plane, so the intersection of the disk's plane with the trajectory plane is
\begin{equation}
    y'\cos i\cos\varphi=z'\sin i.
\end{equation}
Using the equations for the trajectory $y'=r\sin \phi$ and $z'=r\cos\phi$, where $r=1/u$ and $\phi$ are calculated using (\ref{trajectory equation}), we find the angle at which the intersection takes place to be
\begin{equation}
    \label{intersection angle}
    \phi=\arctan{\para{\frac{\tan i}{\cos \varphi}}} + \kappa\pi \qquad k \in \mathbb{Z}.
\end{equation}
The above equation gives a multivalued solution for $\phi$, which can be understood in the following way. The solutions $0 < \phi < \pi$ represent the case when the photon intersects the disk in the upper half $y'Oz$ plane. When $\pi < \phi < 2\pi$, the photon trajectory is bent downwards and intersects the disk in the lower half plane. Values of $\phi > 2\pi$ indicate photons that loop around the black hole. This is possible since the accretion disk has a finite inward extent (we generally considered an extent down to $3R_S$), hence impact parameters between $b_C$ and $3R_S$ may correspond to trajectories that intersect with the disk only after looping around the black hole, as illustrated e.g. in panels (b) and (d) of Fig.~\ref{fig:Trajectories}. Thus,
by giving integer values to $\kappa$, we can take into account the intersection of photons that loop around the black hole one or more times, as numerically these intersections will take place at angles $\phi>2\pi$.

\subsection{Head-on view}\label{sec:bh_image:head-on}

We first consider the case when the black hole accretion disk is contained in the $xOz$ plane, corresponding to a vanishing inclination angle, $i = 0$. The difficulty of simulating the visuals is greatly reduced due to the spherical symmetry. From Eq.~\eqref{intersection angle}, we can see that $\phi=\kappa \pi$ and there is no dependence on the angle $\varphi$ of the rotation matrix in Eq.~\eqref{eq:varphi_def}. This means that photons coming at any angle $\varphi$ that lie on a circle with the same impact parameter $b$ will intersect the disk at the same point.
Thus, it is enough to only check the intersection of photons coming on one axis (for simplicity, the $y$ axis), calculate the temperature at the respective intersection point using Eq.~\eqref{temperature distribution}, create the color for the plot based on the blackbody radiation spectrum, take into account the gravitational redshift, and plot a ring with radius equal to the impact parameter $b$ with the resulting color.

In creating the images, an interval of equidistant values was considered for the impact parameter b (in multiples of $R_S$). The angle at which there can exist an intersection is, according to Eq.~\eqref{intersection angle}, $\phi=\kappa \pi$. After the trajectory is computed, values from $0$ up to $\left \lfloor{ \phi_\infty /\pi }\right \rfloor $ are given to $\kappa$. If an intersection with the disk is found, the corresponding value of $b$ and the radius $R_{\rm intersection}$ on the disk at which the intersection takes place are written in a file to be used in the image processing.

\begin{figure}[t]
    \centering
\begin{tabular}{cc}
    \includegraphics[width=0.45\textwidth]    {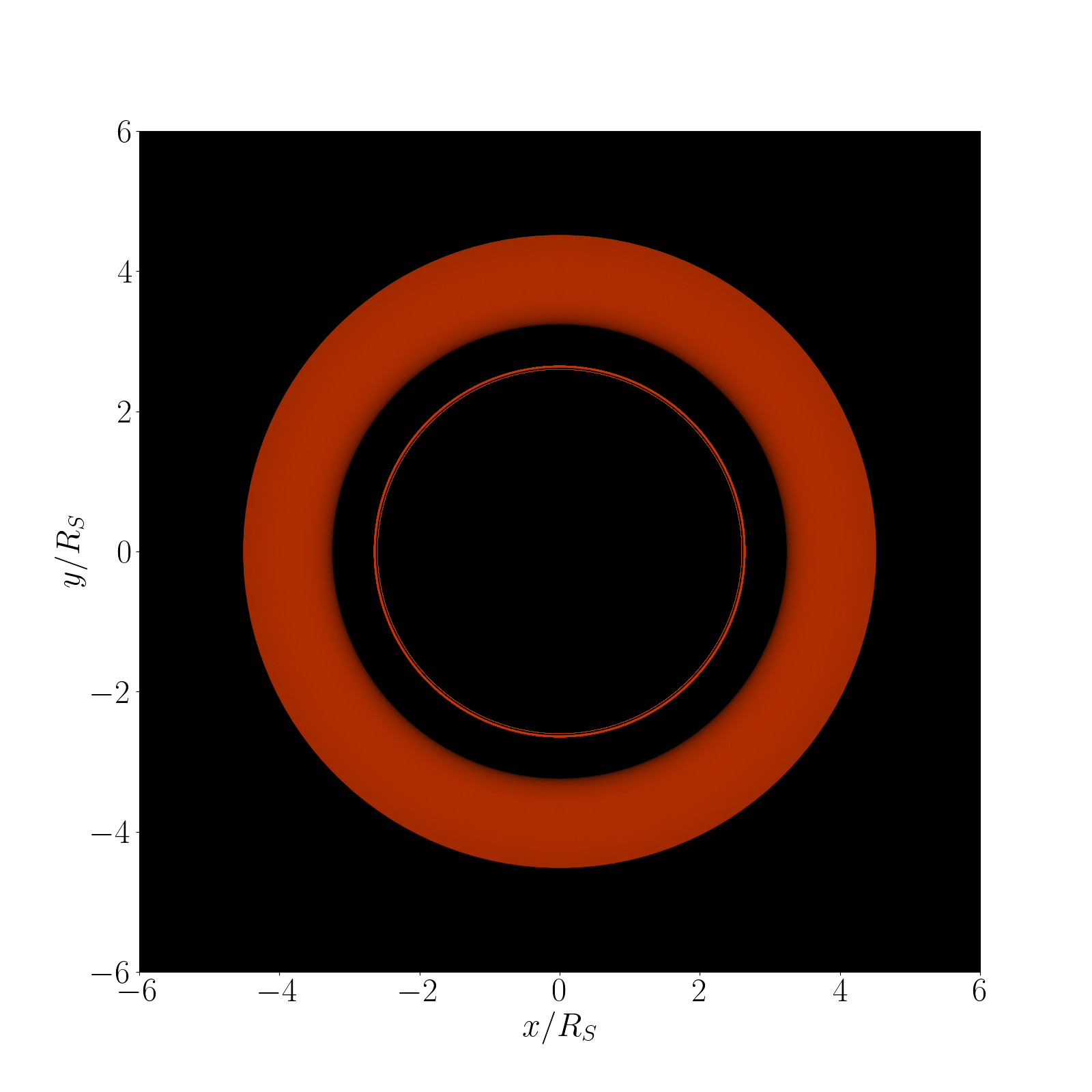} &
    \includegraphics[width=0.45\textwidth]{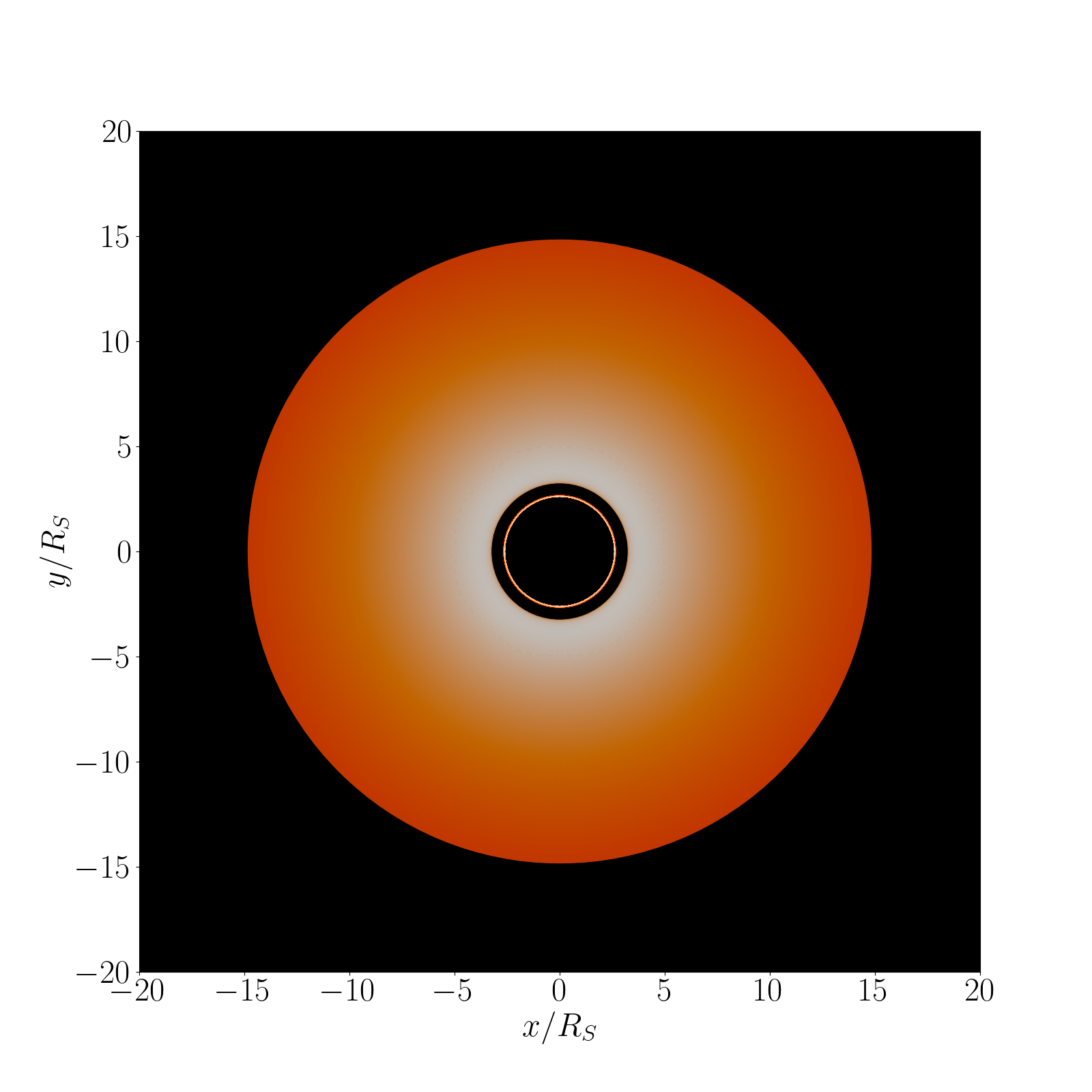} \vspace{-10pt}
\end{tabular}
    \caption{Head-on view of black holes with disk  sizes (left) $3 R_S \to 7R_S$ and (right) $3R_S \to 100R_S$, obtained as described in Sec.~\ref{sec:bh_image:head-on}. It can be observed that the image is smaller than the disk due to the gravitational lensing effect.
    \label{fig:Head-on}
    }
\end{figure}

We determine the color based on Tanner Helland's blackbody approximation algorithm \cite{bib:Tanner Helland}. Although the algorithm was not built for scientific use, it is sufficient for our purpose. The algorithm takes a temperature in Kelvin, given in our case by Eq.~\eqref{temperature distribution}, and clamps it in the interval [1000 K, 40000 K]. Each color is calculated based on whether the temperature corresponds to warmer colors or cooler colors (we took the splitting value to be $6600$ K, the same as in the sample code of Ref.~\cite{bib:Tanner Helland}). A logarithmic curve based on the value of $T$ is used to find the corresponding value between $0$ and $255$ for each RGB color channel.
The image is created by taking the values of $b$ and $R_{\rm intersection}$ from the saved data file and by coloring rings of radius $b$ based on Tanner Helland's algorithm \cite{bib:Tanner Helland}, where $R_{\rm intersection}$ is used in equation \eqref{temperature distribution} to compute the temperature of the disk at the intersection.
Simulated images of head-on black holes are presented in Fig.~\ref{fig:Head-on}.

The thin inner ring in both panels of Fig.~\ref{fig:Head-on} represents the photon sphere at the photon capture radius, $r = 1.5 R_S$ \cite{Luminet:1979nyg}. Figure~\ref{fig:Head-on} (b) depicts a bigger, energetic accretion disk, which can be found, e.g., around quasars. Gravitational redshift/blueshift effects are more evident in this case as the photons closer to $3 R_S$ are subject to a higher gravitational pull, in consequence suffering a bigger shift than the photons at the edge of the disk.

\subsection{Inclined disc view}\label{sec:bh_image:incline}

\begin{figure}[t]
\centering
\begin{tabular}{cc}
 \includegraphics[width=.45\textwidth]{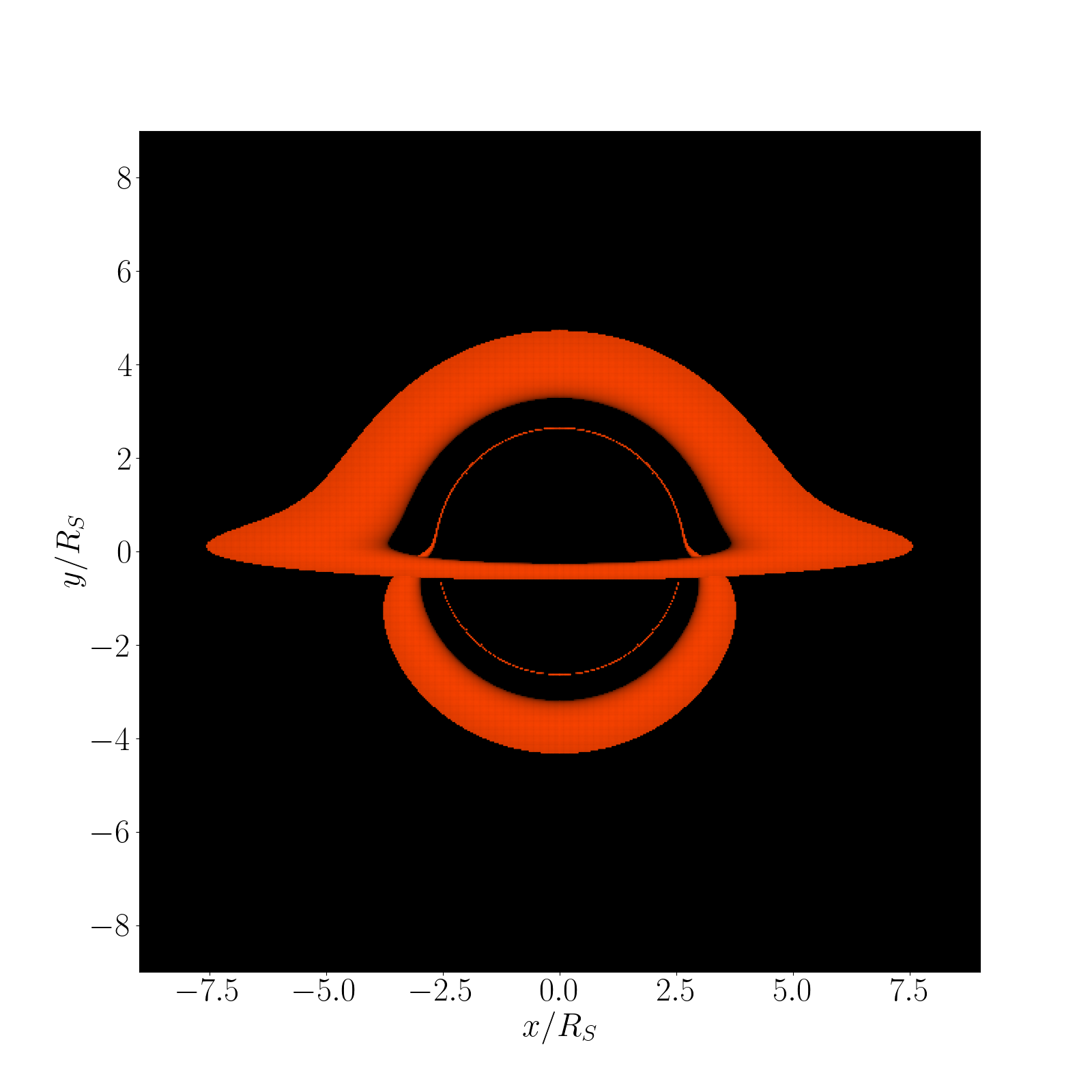} &
 \includegraphics[width=.45\textwidth]{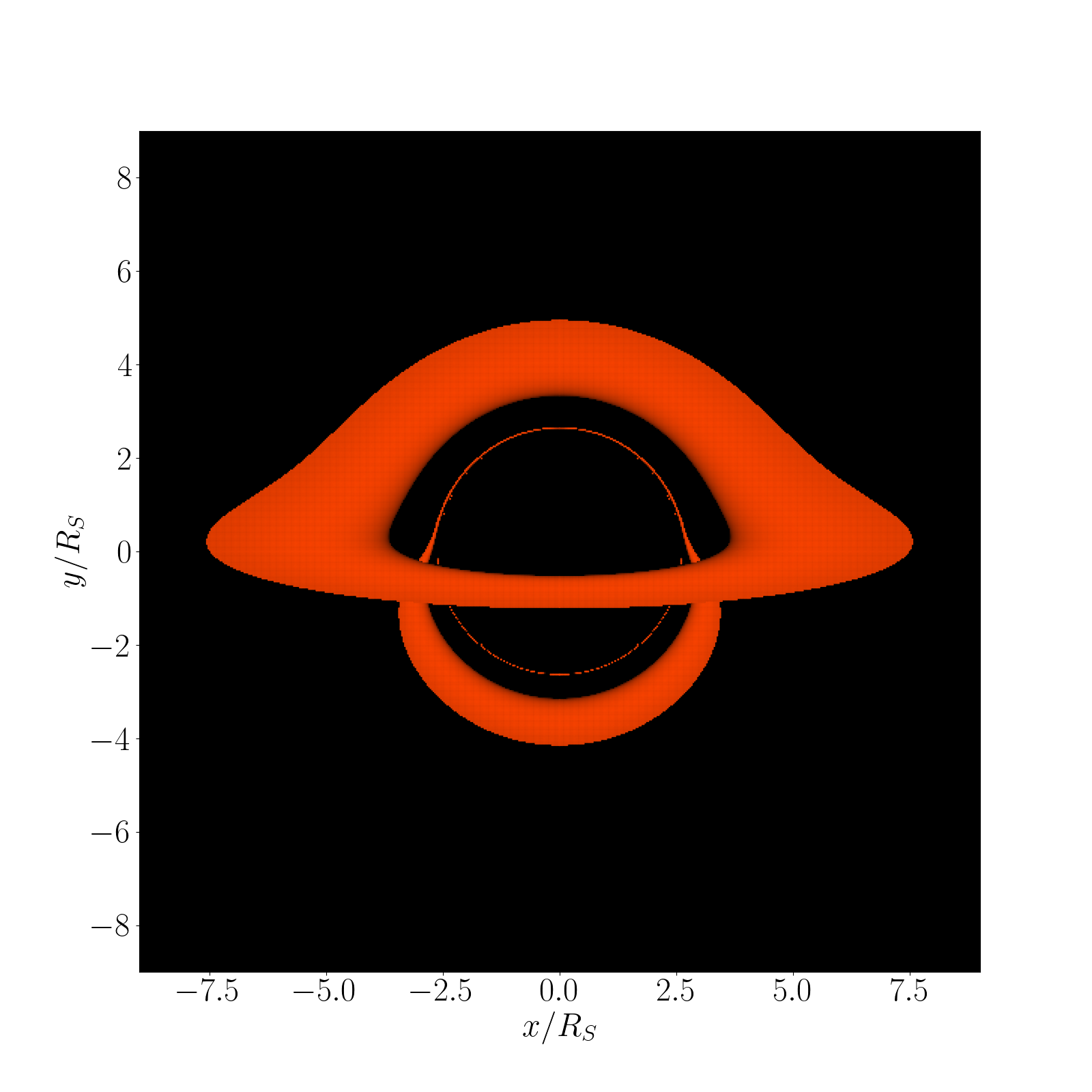}  \vspace{-42pt} \\
 {\scriptsize \color{white} (a) $i = -5 ^ \circ$} &
 {\scriptsize \color{white} (b) $i = -10 ^ \circ$} \\
 \includegraphics[width=.45\textwidth]{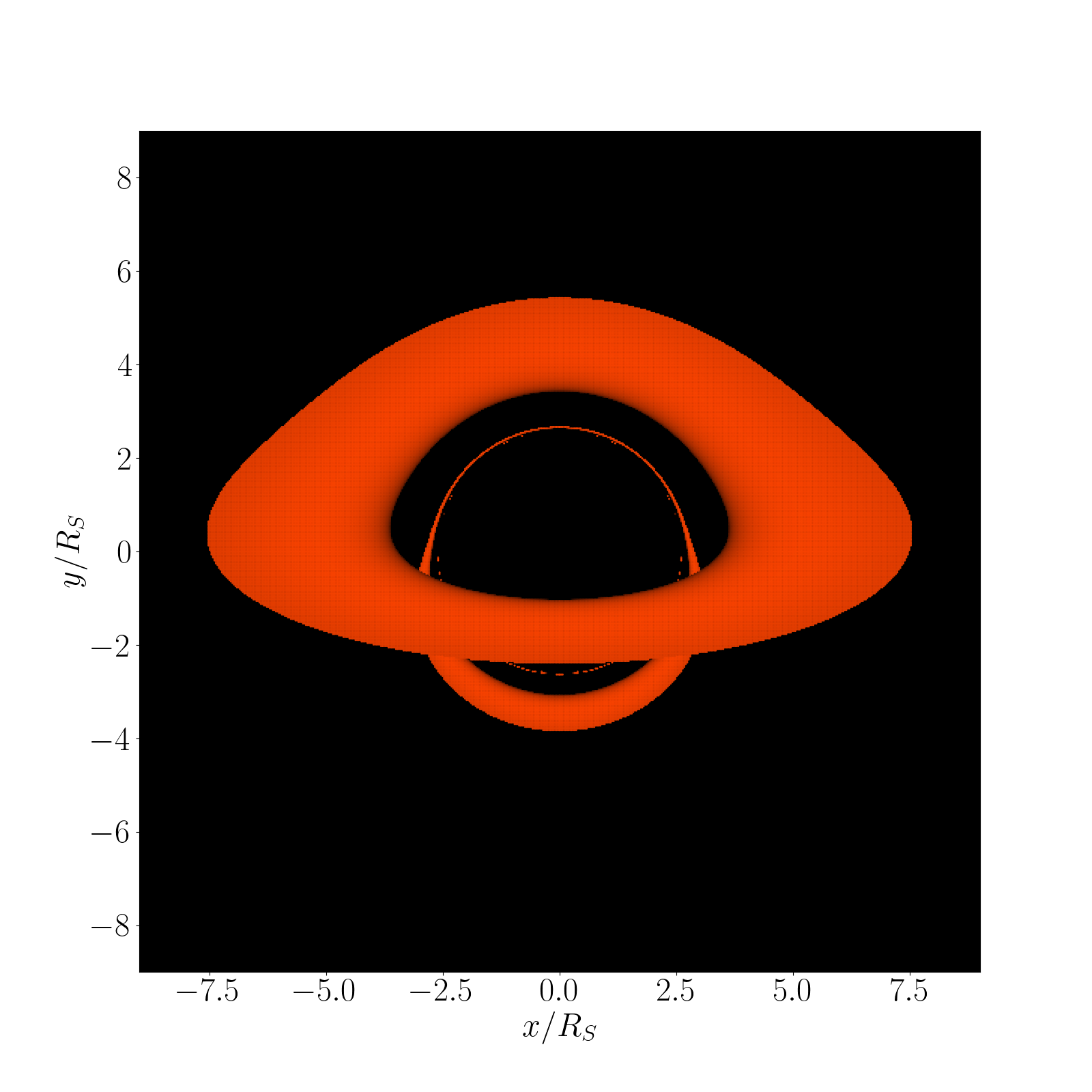} &
 \includegraphics[width=.45\textwidth]{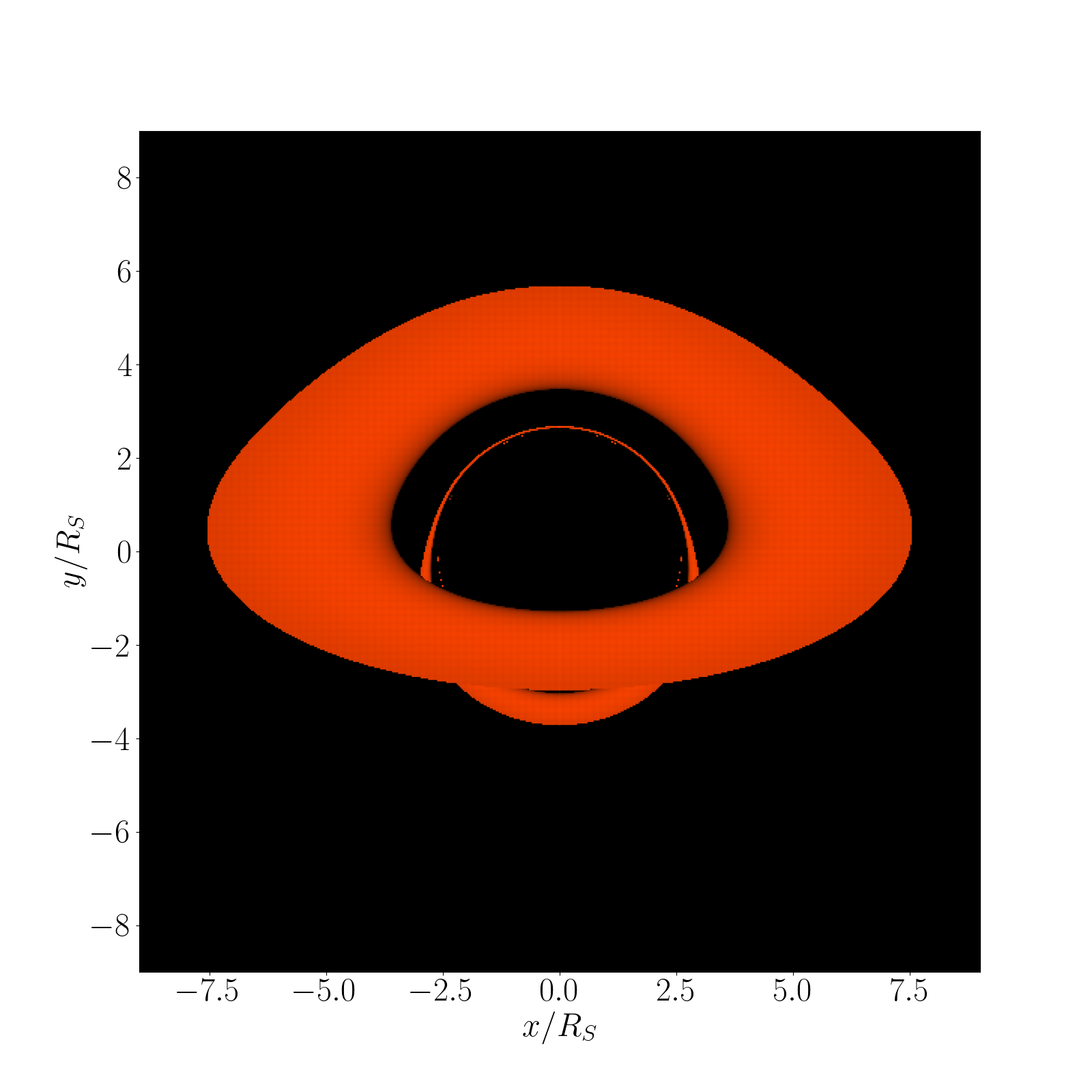}  \vspace{-42pt} \\
 {\scriptsize \color{white} (c) $i = -20 ^ \circ$} &
 {\scriptsize \color{white} (d) $i = -25 ^ \circ$} \vspace{10pt}
\end{tabular}
    \caption{Black holes with inclined accretion disks. All disks span from $3R_S$ to $7R_S$. Sample points: $500 \times 500$.}
    \label{fig:Inclined}
\end{figure}

If the disk makes an angle $i\neq 0$ with the incident axis (observer $\to$ black hole), the intersection angle for a photon with the disk is given by Eq.~\eqref{intersection angle}. We therefore cannot adopt the same strategy employed in Subsec.~\ref{sec:bh_image:head-on}. In this case, a grid of points each corresponding to a test photon must be considered. In sampling the grid, the problem of photons coming from the bottom ($y<0$) is equivalent to that of photons coming from the top towards a black hole with the accretion disk at an angle $i'=-i$ (practically flipping the setup around the incident axis).

In sampling grid points, only the upper-right and lower-right quadrants of the viewing plane were considered in calculations because the left side will behave the same as the right side due to the spherical symmetry of the system. For both quadrants, an equidistant interval of points was considered for both $x$ and $y$ (measured from the origin of the viewing plane) between 0 and a set upper boundary (a value larger than the considered radius of the accretion disk). The tilt of the photon's plane and the $y$ axis (angle $\varphi$ in equation \eqref{intersection angle}) is calculated as $\varphi=\arctan{x/y}$ and for each trajectory. As for $\kappa$, values from 0 up to $\left \lfloor{ \phi_\infty /\pi}\right \rfloor $ are given. If an intersection is found, the corresponding values for $x$ and $y$ and the radius $R_{\rm intersection}$ on the disk at which the intersection takes place are written in a file.
The image is created by adding a small colored dot at the corresponding point $(x,y)$ with the colour given by Tanner Helland's blackbody approximation algorithm \cite{bib:Tanner Helland}.
Black holes with accretion disks at different inclination angles are shown in Fig.~\ref{fig:Inclined}.

The extreme bending of the spacetime near the black hole leads to photons going around the black hole, making it possible to see the back of the accretion disk as a halo over and under the black hole. The photons which go around the photon sphere at $r = 1.5 R_S$ can also be seen in the inner thin disk, localized around the circle of radius $b_C = \frac{3 \sqrt{3}}{2} R_S$.

As the inclination angle $i$ approaches 0, the image becomes more warped around the black hole, eventually arriving at the head-on view (Figure \ref{fig:Head-on}) in which the disk is invisible from the front but the backside can be seen due to the bending of the photon trajectories.

\subsection{Rotating accretion disk}\label{sec:bh_image:rot}

A simple way in which the rotation of the accretion disk can be modelled is to consider the Keplerian circular orbit angular velocity \cite{Luminet:1979nyg,bib:Accretion} for each point on the accretion disk
\begin{equation}
    \label{keplerian angular velocity}
    \omega=\sqrt{\frac{GM}{r^3}}.
\end{equation}
The redshift will now not be solely due to the gravitational effects, but also because of relativistic Doppler effects. In this case, the ratio of the received and emitted frequencies is \cite{bib:Hobson,Luminet:1979nyg}
\begin{equation}
    \label{general redshift aux}
    \frac{\nu_R}{\nu_E}=\frac{p_\mu u_R^\mu(R)}{p_\mu u^\mu_E(E)},
\end{equation}
where $p_\mu$ is the covariant 4-momentum of the photon, $u^\mu_R$ is the 4-velocity of the receiver and $u^\mu_E$ is the 4-velocity of the emitter.

The observer is considered to be at rest in the far-field limit, so $u^\mu_R=(c,0,0,0)$. The 4-velocity of the emitter is $u^\mu_E=(u^t,0,0,u^\phi)$. We can establish a relation between $u^t$ and $u^\phi$ by considering
\begin{equation}
    \omega =\frac{d \phi}{dt}=\frac{d\phi/d\lambda}{dt/d\lambda}=\frac{u^\phi}{u^t} \Longrightarrow u^\phi=\omega u^t.
    \label{eq:uphi}
\end{equation}
Therefore, the 4-velocity of the emitter is $u^\mu_E=u^t(1,0,0,\omega)$. On the other hand, from \eqref{physical interpretation of k and h} and the definition of the impact parameter, we can write
\begin{equation}
    \frac{p_\phi}{p_t}=-\frac{b_\phi}{c},
    \label{eq:pphi}
\end{equation}
where we used the notation $b_\phi$ to indicate that this quantity refers to the projection of the axial impact parameter $b$ (which in our algorithm is computed as the distance $b=\sqrt{x^2+y^2}$ from the center of our motion) on the axis of rotation \cite{Luminet:1979nyg,Cunningham:1975zz}.
Substituting Eqs.~\eqref{eq:uphi} and \eqref{eq:pphi} into \eqref{general redshift aux} gives the following expression for the redshift:
\begin{equation}
    \frac{\nu_R}{\nu_E}=\frac{c}{u^t(1\pm b_\phi\omega/c)}.
\end{equation}
where the plus sign corresponds to recession of the disk and the minus sign corresponds to the approaching side \cite{Luminet:1979nyg,Cunningham:1975zz}.
An expression for $u^t$ can be derived from the invariant $u_\mu u^\mu =g_{\mu\nu} u^\mu u^\nu=c^2$. Summing over the indices and simplifying, we arrive at
\begin{equation}
    u^t=\frac{c}{\sqrt{g_{tt}+\omega ^2 g_{\phi\phi}}}.
\end{equation}
Thus the total redshift is
\begin{equation}
    \label{general redshift}
    \frac{\nu_R}{\nu_E}=\frac {\sqrt{g_{tt}+\omega ^2 g_{\phi\phi}} } {1\pm b_\phi\omega / c},
\end{equation}
which reduces to \eqref{redshift} when $\omega =0$.

\begin{figure}[t]
\begin{tabular}{cc}
 \includegraphics[width=.45\textwidth]{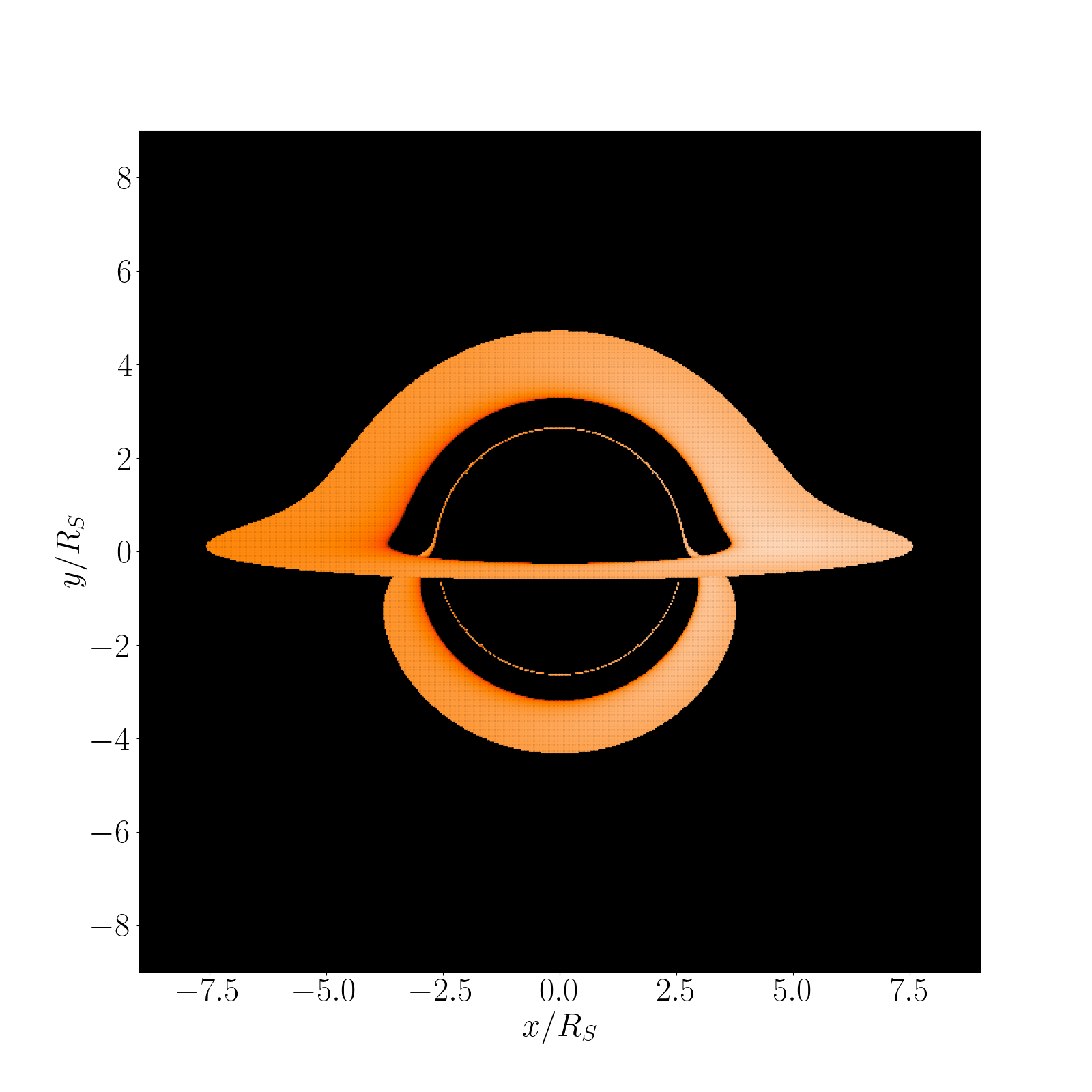} &
 \includegraphics[width=.45\textwidth]{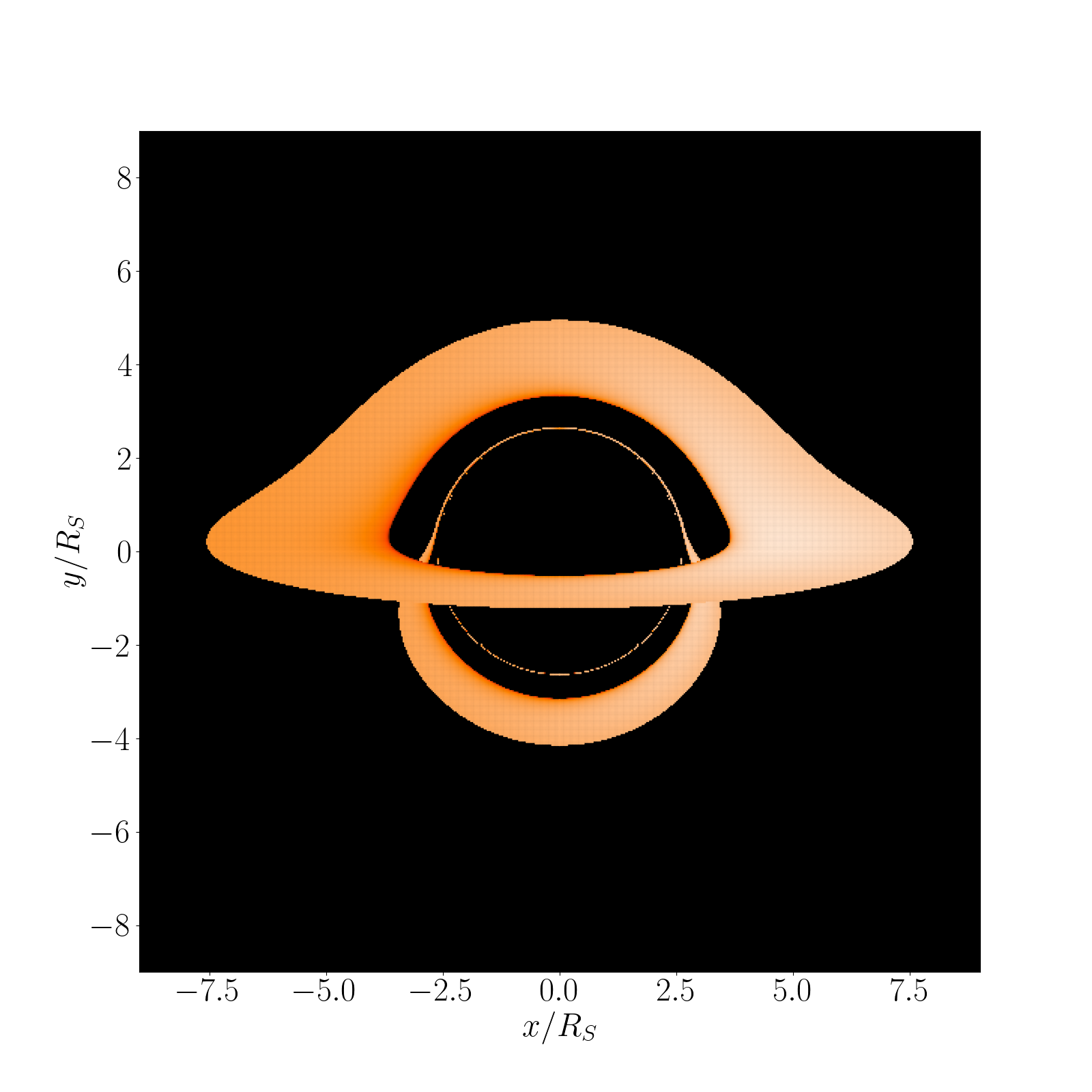} \vspace{-42pt} \\
 {\scriptsize \color{white} (a) $i = -5 ^ \circ$} & {\scriptsize \color{white} (b) $i = -10 ^ \circ$}  \\
 \includegraphics[width=.45\textwidth]{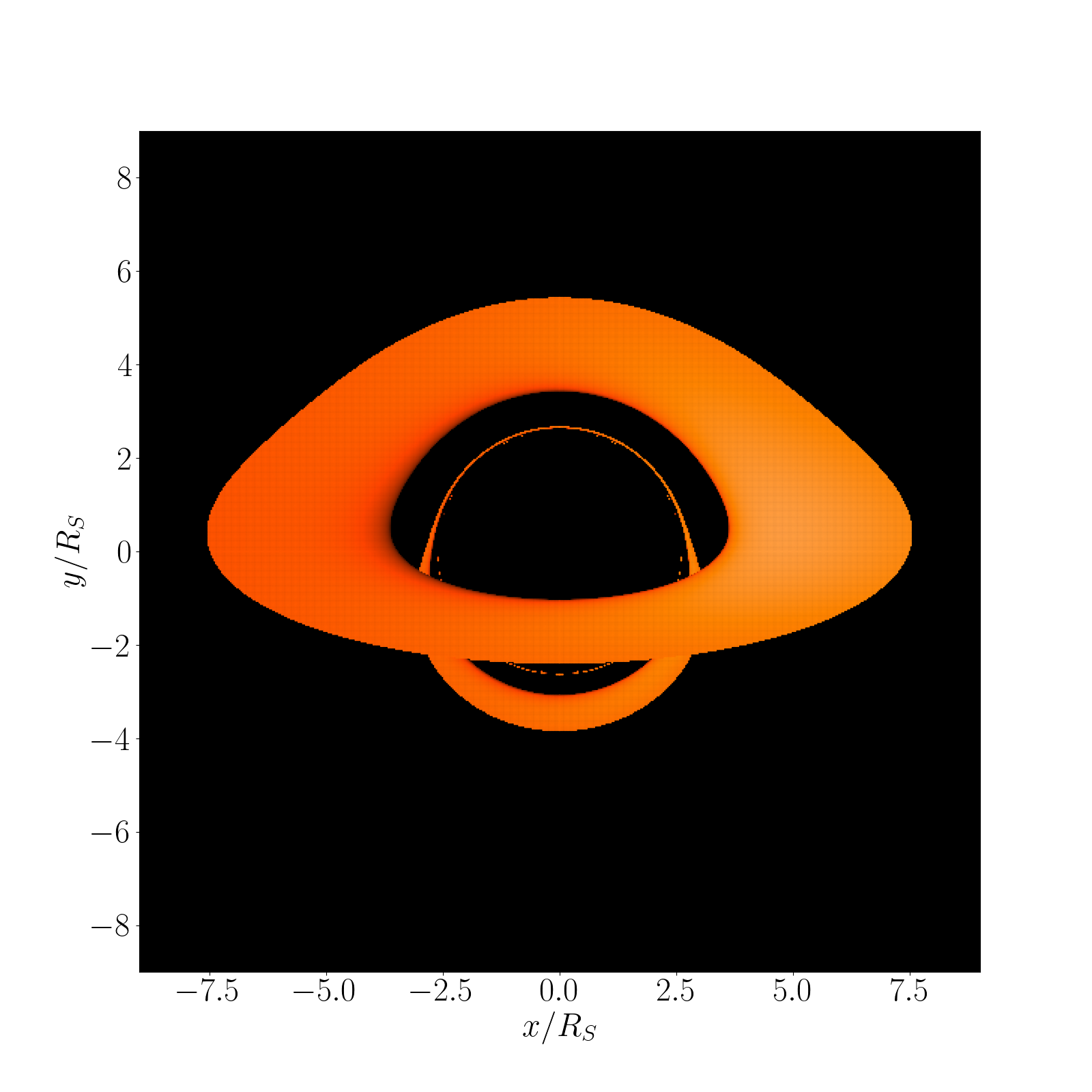} &
 \includegraphics[width=.45\textwidth]{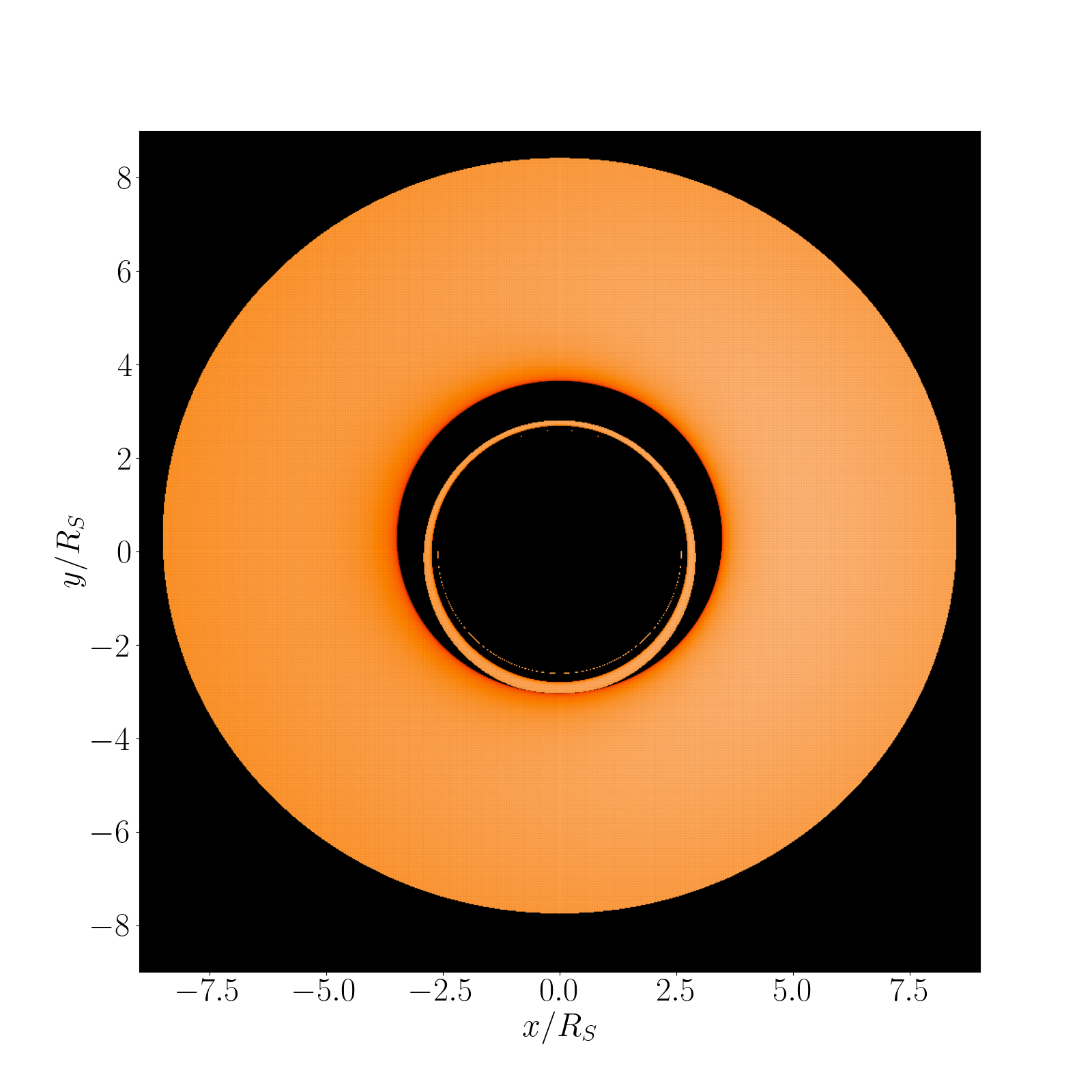} \vspace{-42pt} \\
 {\scriptsize \color{white} (c) $i = -20 ^ \circ$} & {\scriptsize \color{white} (d) $i = -70 ^ \circ$} \vspace{10pt}
\end{tabular}
    \caption{Black holes with rotating accretion disks at different inclination angles and temperature profiles: (a) $-5 ^ \circ$, 500x500 sample points; (b) $-10 ^ \circ$, 500x500 sample points; (c) $-20 ^ \circ$, 500x500 sample points; (d) $-70 ^ \circ$. 1000x1000 sample points  }
    \label{fig:Rotating disks}
\end{figure}

In generating the black hole images, the same strategy as in Subsection \ref{sec:bh_image:incline} was used. The only difference appears at image processing, where we use \eqref{general redshift} instead of \eqref{redshift} for the redshift of the photons.
Images with disks at different temperature profiles, that can be adjusted by varying the free parameter $\alpha$ in \eqref{temperature distribution}, are displayed in Fig.~\ref{fig:Rotating disks}.

We can see that the simulated images presented in Fig.~\ref{fig:Rotating disks} correctly predict a blueshift of the approaching side of the disk (to the right of the figure) and a redshift of the receding side of the disk (to the left of the figure). Note that the disk is rotating in the clockwise direction with respect to its symmetry axis.
The parameters employed for the black hole of Fig.~\ref{fig:Rotating disks}(d) correspond to those of the M87 black hole \cite{bib:EHT2019}, namely an inclination angle of $i \simeq 17^{\circ} - 90^{\circ} = -73^{\circ}$ (we used $-70^{\circ}$) and an accretion disk of extent $d \simeq 5R_S $ \cite{bib:EHT2019}.

\begin{figure}[H]
    \centering
\begin{tabular}{cc}
    \includegraphics[height=200pt]{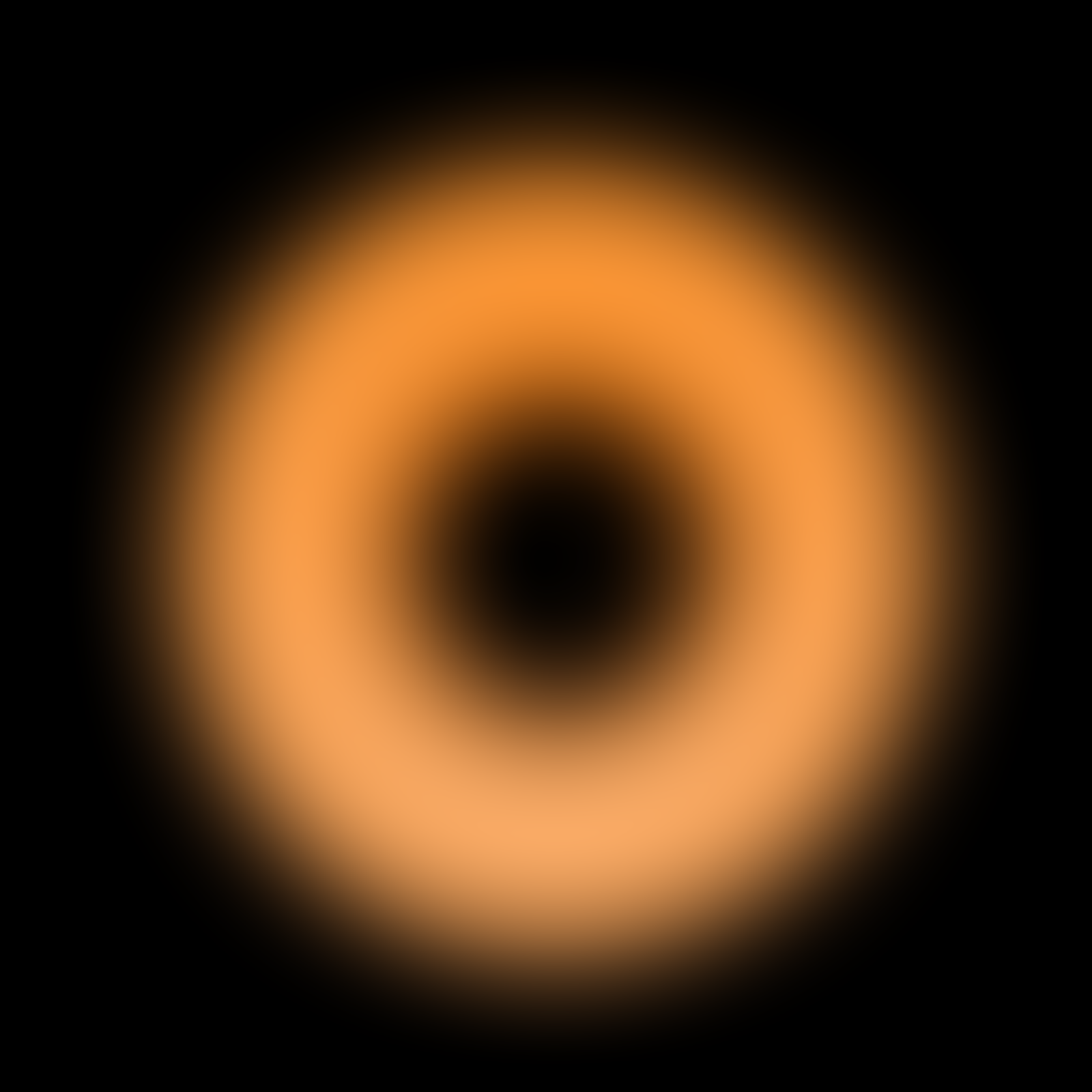} &
    \includegraphics[height=200pt]{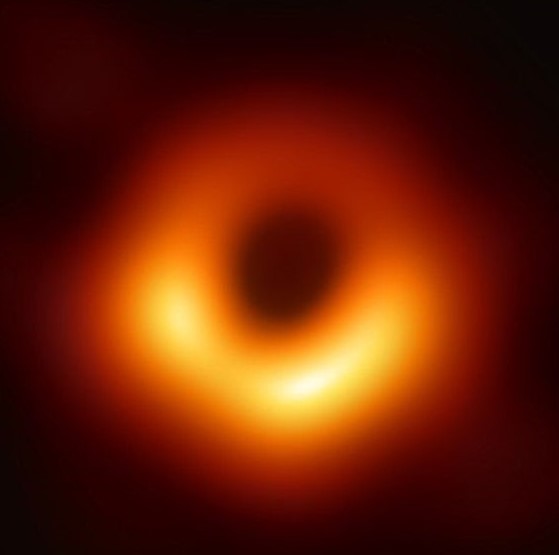}
\end{tabular}
    \caption{Comparison of (left) our simulated image, obtained by blurring Fig.~\ref{fig:Rotating disks} (d), to (right) the image of the black hole at the center of M87 captured by the Event Horizon Telescope~\cite{bib:EHT2019}.
    \label{fig:M87 comparison}
    }
\end{figure}

In order to compare our simulation results with the image reported by the EHT collaboration, shown in Fig.~\ref{fig:M87 comparison} (right panel), we employed a standard blurring algorithm on Fig.~\ref{fig:Rotating disks}(d), resulting in the image shown in the left panel of Fig.~\ref{fig:M87 comparison}.

We can see from Fig~\ref{fig:M87 comparison} that our simulation catches the general shape of the accretion disk around the black hole. However, the model does not accurately predict the radiation spectrum emitted, the shift in photon wavelength being much weaker in our model. These discrepancies arise from our consideration of a Schwarzschild black hole instead of a Kerr black hole and due to the simplistic model of the accretion disk. The dimensionless spin parameter $a = J c / GM^2$ of the M87 black hole is poorly constrained, however values of $|a|$ as high as $0.94$ (close to the maximal limit $a = 1$) are considered in Ref.~\cite{EHT:2025dua}, corresponding to a rapidly rotating Kerr black hole. More accurate descriptions of the accretion disk involve general-relativistic magnetohydrodynamical (GRMHD) simulations (see, e.g., Refs.~\cite{Porth:2016rfi,Prather:2024hsu}), which are beyond the scope of our simplistic considerations.

\section{Conclusions} \label{sec:conc}

In this paper, we considered the problem of black hole imaging in the Schwarzschild geometry using the photon geodesic equations \eqref{eq:null} as the basis of the ray tracing algorithm. The approach of using a Runge-Kutta fourth order algorithm proves to be highly precise in determining deflection angles, as seen in Fig.~\ref{fig:Convergence plots}.

The weak field approximation in Eq.~\eqref{weak field angle} was also tested against our simulations. As seen from Fig.~\ref{fig:deflection angle}, the approximation matches the lower part of the graph (up to deflection angles of around $\delta \phi \simeq 3^\circ$ ). Considering the weak field approximation without simplifying the trigonometric functions using the small angle expansions proves to be a worse approximation (Fig.~\ref{fig:deflection angle}) than the one given by \eqref{weak field angle} .

Simulated images of the black holes, presented in Fig.~\ref{fig:Head-on}, Fig.~\ref{fig:Inclined} and Fig.~\ref{fig:Rotating disks}, showcase expected phenomena such as strong gravitational lensing of the photon sphere in the region close to the black hole, redshift of the emitted photons due to gravitational effects and due to rotation of the accretion disk (Fig.~\ref{fig:Rotating disks}) and manages to capture the general size of the black hole shadow and disk as viewed by the observer (Fig.~\ref{fig:M87 comparison}).

Due to our consideration of the Schwarzschild metric, we fail to predict the asymmetry of the image and the redshift effects due to the rotation of the black hole. The Kerr metric should be used in a more accurate description of the system~\cite{bib:EHT2019,EHT:2025dua}. In addition, due to the simplifications employed in the thin accretion disk model, we fail to predict the powerful brightness and spectral differences due to the strong magnetic fields and turbulent nature of the accretion disk shown in Fig~\ref{fig:M87 comparison}. We point out the necessity of employing GRMHD simulations of the matter surrounding the accretion disk for a more accurate model (e.g., Refs.~\cite{Porth:2016rfi,Prather:2024hsu}).

{\bf Acknowledgments.}

The authors are grateful to Dr. Nyx Shiva for insightful discussions.

\bibliographystyle{plain}

\end{document}